\documentclass[journal]{IEEEtran}
\usepackage{epsfig,latexsym}
\usepackage{float}
\usepackage{indentfirst}
\usepackage{amsmath}
\usepackage{amssymb}
\usepackage{times}
\usepackage{subfigure}
\usepackage{psfrag}
\usepackage{cite}
\usepackage{color}
\usepackage{algorithm}
\usepackage{setspace}
\usepackage{mathrsfs}
\usepackage{stfloats}
\usepackage{graphicx}

\begin{document}
\begin{sloppypar}
\title{\huge{Joint Optimization of Resource Allocation, Phase Shift and UAV Trajectory for Energy-Efficient RIS-Assisted UAV-Enabled MEC Systems}}

\author{Xintong Qin, Zhengyu Song, Tianwei Hou, Wenjuan Yu, Jun Wang, and Xin Sun

\thanks{X. Qin, Z. Song, T. Hou, J. Wang, and X. Sun are with the School of Electronic and Information Engineering, Beijing Jiaotong University, Beijing 100044, China (email: 20111046@bjtu.edu.cn, songzy@bjtu.edu.cn, twhou@bjtu.edu.cn, wangjun1@bjtu.edu.cn, xsun@bjtu.edu.cn).}
\thanks{W. Yu is with the School of Computing and Communications, InfoLab21, Lancaster University, Lancaster LA1 4WA, U.K. (e-mail:  w.yu8@lancaster.ac.uk).}
}

\markboth{}
{Shell \MakeLowercase{\textit{et al.}}: Bare Demo of IEEEtran.cls for Journals}

\maketitle

\begin{abstract}
The unmanned aerial vehicle (UAV) enabled mobile edge computing (MEC) has been deemed a promising paradigm to provide ubiquitous communication and computing services for the Internet of Things (IoT). Besides, by intelligently reflecting the received signals, the reconfigurable intelligent surface (RIS) can significantly improve the propagation environment and further enhance the service quality of the UAV-enabled MEC. Motivated by this vision, in this paper, we consider both the amount of completed task bits and the energy consumption to maximize the energy efficiency of the RIS-assisted UAV-enabled MEC systems, where the bit allocation, transmit power, phase shift, and UAV trajectory are jointly optimized by an iterative algorithm with a double-loop structure based on the Dinkelbach’s method and block coordinate decent (BCD) technique. Simulation results demonstrate that: 1) with the deployment of RIS, our proposed algorithm can achieve higher energy efficiency than baseline schemes while satisfying the task tolerance latency; 2) the energy efficiency first increases and then decreases with the increase of the mission period and the total amount of task-input bits of IoT devices; 3) when the CPU cycles required for computing 1-bit of task-input data becomes larger, more task bits will be offloaded to the UAV while the energy efficiency will be decreased.

\end{abstract}


\section{Introduction}
In recent years, with the rapid development of Internet of Things (IoT), the number of IoT devices are dramatically increasing, which spurs the emerging of more and more novel intelligent applications, such as augmented reality, virtual reality, face recognition, and so forth \cite{YMao2017}. The tasks generated by these applications usually demand for higher computation capacity and lower latency. However, it is not reliable to depend on the IoT devices themselves to satisfy such computation demands, due to the limited computation capacity and energy budget. In this context, mobile edge computing (MEC) appears as a promising paradigm to support IoT devices' computation-intensive and latency-critical tasks \cite{FWang2018}. By deploying the MEC server at a wireless access point (AP) or base station (BS), it provides offloading opportunities for IoT devices, and thus can help IoT devices compute tasks and save energy. Despite the benefits of infrastructure-based MEC systems, the fixed location of MEC server restricts the coverage and the service will not be restored in a short time if the infrastructure is destroyed. Recently, the unmanned aerial vehicles (UAVs) characterized by their mobility, flexibility, and maneuverability have drawn considerable
attentions. It is envisioned that the UAV-enabled MEC can be widely deployed to provide ubiquitous communication and computing supports for IoT devices, which has received growing popularity \cite{YZeng2019}.

By equipping an MEC server, the UAV can provide computing and communicating services for IoT devices. Thanks to the inherent advantages such as flexible deployment and high mobility, the benefits of integrating the UAV into MEC are multifold. For example, the UAV can be rapidly deployed on the hotspots to cooperate with ground base stations, and thus enhance the computing capacity of the network to tackle the surge of computation demands during rush hours \cite{ZYu2020}. In addition, due to the high altitude of UAV, the probability of line-of-sight (LoS) links can be improved, which is beneficial for IoT devices’ energy saving when they perform task offloading \cite{HGuo2020}. Moreover, the flexible trajectory of UAV brings an additional degree of freedom for communication performance enhancement \cite{QHu2019}. By optimizing the trajectory, the UAV can move closer toward the IoT devices to obtain better channel conditions and achieve lower energy consumption in task offloading \cite{XQin2021}. However, the UAV-enabled MEC systems still face many challenges. For example, the direct links among the UAV and IoT devices may be blocked by buildings in urban areas, which heavily degrades the channel conditions. Besides, due to the limited energy budget of UAV, it is essential to jointly optimize the UAV trajectory and computing energy consumption to achieve higher energy efficiency.

In order to fully reap the benefits of integrating the UAV into MEC and further improve the IoT devices’ task offloading performance, an emerging paradigm called reconfigurable intelligent surface (RIS) has drawn great attentions \cite{CPan2020}. The RIS is a planar array consisting of a large number of reflecting elements \cite{YLiu2021}. Through adjusting the phase shift of RIS element, the concatenated virtual LoS link is formed between the transmitter and receiver. Thus, the reflected signals can be combined coherently to improve the received signal power \cite{QWu2020}. In addition, compared with conventional relay systems, the RIS passively reflects signals without power amplification, which is more environment-friendly and can improve the energy efficiency of the whole system \cite{QWu20202}. Moreover, the RIS can be flexibly deployed on various structures, such as the building facades, roadside billboards, and indoor walls, which makes it trivial to integrate the RIS into existing wireless networks \cite{SGong2020}.

Despite the above-mentioned advantages, there are still some urgent challenges to be addressed before RIS can be harmoniously integrated into the UAV-enabled MEC systems. For example, the phase shift optimization for RIS provides an additional degree of freedom (DoF) to enhance the performance of UAV-enabled MEC systems. However, the communication and computing resource allocation for MEC is closely coupled with the phase shift optimization, which results in a non-convex optimization problem. Besides, although the UAV trajectory design has been intensively studied in UAV-assisted networks without RIS, the proposed algorithms cannot be directly applied to optimize the UAV trajectory with the participation of RIS. This is because the angle of departure of the signal from the RIS to the UAV varies with the positions of UAV. Thus, the UAV trajectory design is coupled with the phase shift optimization. When jointly considering the interaction of resource allocation, phase shift and UAV trajectory, the resulting optimization problem is highly non-convex and non-trivial to derive a globally optimal solution. Additionally, the deployment of RIS to a certain extent facilitates the task offloading of IoT devices, but when a large amount of task bits is offloaded to the UAV-mounted MEC server, it will threaten the system's operating time since the energy storage of the UAV is heavily limited. Hence, considering this contradiction between the RIS and the UAV-enabled MEC, it is of great importance to investigate the energy efficiency by simultaneously paying attention to the energy consumption and the amount of completed task bits. In recent years, intensive efforts have been devoted to studying the RIS, UAV-enabled communications, and the fusion of UAV and MEC systems. However, there are still few studies focusing on the optimization of MEC systems assisted by both the UAV and RIS. Therefore, to bridge this research gap and tackle the aforementioned challenges, we investigate the RIS-assisted UAV-enabled MEC systems in this paper with the objective to maximize the energy efficiency, by jointly optimizing the task bit allocation between IoT devices and the MEC server, transmit power of IoT devices, phase shift of RIS, and the UAV trajectory. The main contributions of this paper can be summarized as follows.

1) We first establish a novel RIS-assisted UAV-enabled MEC framework to explore the potential benefits of RIS in UAV-enabled MEC systems, where the RIS is deployed on the surrounding building wall to assist IoT devices' task offloading, and the UAV equipped with an MEC server provides offloading opportunities and computing services for multiple IoT devices. During the task offloading, both the direct links from IoT devices to the UAV and the reflecting links via the RIS are considered. Besides, partial offloading scheme is applied in the IoT devices and all devices access the UAV by non-orthogonal multiple access (NOMA) protocol.

2) By taking both the amount of completed task bits and energy consumption into consideration, an energy efficiency maximization problem is formulated for the RIS-assisted UAV-enabled MEC system. Due to the fractional structure and highly-coupled variables of the formulated problem, an iterative algorithm with a double-loop structure is proposed, where the outer loop is used to update the energy efficiency based on the Dinkelbach's method, and the inner loop is decomposed into three subproblems to iteratively optimize the bit allocation, transmit power, phase shift, and UAV trajectory. For the three subproblems in the inner loop, the Langrange dual method is utilized to solve the bit allocation and transmit power optimization problem; the phase shift of RIS is optimized by the difference of convex functions (DC) programming; and the successive convex approximation (SCA) technique is exploited to tackle the UAV trajectory optimization.

3) Extensive simulation results verify that with the deployment of RIS, our proposed algorithm can achieve higher energy efficiency compared to the schemes with random phase, without trajectory optimization, without RIS, and the full offloading scheme. Meanwhile, the UAV tends to fly closer to the RIS in order to acquire a better channel condition to improve the energy efficiency. Interestingly, it can be found that with the increase of the mission period, the energy efficiency first increases and then decreases since the flying energy consumption of UAV continues to increase and
dominates the total energy consumption. Besides, with the increase of the total amount of task-input bits of IoT devices, the energy efficiency also first increases and then decreases. Finally, when the CPU cycles required for computing 1-bit of task-input data becomes larger, the energy consumption for computing will be increased and thus results in the decrease of energy efficiency, while more task bits will be offloaded to the UAV to ensure the tasks can be completed within the mission period.

The rest of the paper is organized as follows. Related works are discussed in Section II. In Section III, we introduce the system model and problem formulation for energy efficiency maximization. Section IV elaborates on the proposed algorithms for solving the formulated problem. Some numerical results are shown in Section V, and conclusions are finally drawn in Section VI.

\section{Related work}
\subsection{UAV-Enabled MEC}
Recently, the UAV-enabled MEC network has been extensively investigated \cite{XZhang2020, JZhang2019,XHu2020,MLi2020, JZhang2020EE, XZhang2019, ZYang2021}. Considering the low-power ground IoT devices and the limited onboard energy storage of UAV, it is of great importance dedicating to the energy consumption minimization problems in the UAV-enabled MEC networks. To this end, X. Zhang \emph{et al.} in \cite{XZhang2020} propose an efficient iterative algorithm to jointly optimize the trajectory design and resource allocation aiming to minimize the weighted sum energy consumption of users and UAVs in the UAV-enabled MEC system. Considering the stochastic offloading scheme, J. Zhang \emph{et al.} in \cite{JZhang2019} adopt the Lyapunov optimization technique to optimize the same objective. Simulation results show that the proposed scheme can save more energy compared with that only considering the energy consumption of users. In addition, compared with the scheme that only minimizes the energy consumption of UAV, more backlogs of the task queues can be processed by tasking both the energy consumption of users and UAV into consideration.

Furthermore, by simultaneously considering the amount of completed task bits and energy consumption, the energy efficiency problems in the UAV-enabled MEC are investigated in \cite{MLi2020, JZhang2020EE, XZhang2019}. To be specific, the energy efficiency of smart mobile devices (SMDs) in a multi-UAV assisted MEC system is investigated in \cite{JZhang2020EE}, where an iterative optimization algorithm based on the  Dinkelbach's method is proposed to tackle the formulated fractional programming problem. Different from \cite{JZhang2020EE}, the energy efficiency of UAV is maximized in \cite{MLi2020} by jointly optimizing the UAV trajectory, the user transmit power, and computation load allocation. Moreover, the node mobility estimation has been adopted in \cite{MLi2020} to design a proactive UAV trajectory when the knowledge of user trajectory is limited, which offer valuable insights on UAV optimal trajectory design for providing on-demand edge computing service for remote IoT nodes.

\subsection{RIS-Assisted Networks}
The general definition of RIS is first given in \cite{RDi2020}. In order to realize the vision of smart radio environment (SRE), RIS is regarded as a key enabler, which has received increasing research attentions from both industry and academia. For example, NTT DOCOMO designs a kind of smart glass, which can dynamically control the response of the impinging radio waves. Another practical example of RIS is called RFocus, which is designed by Massachusetts Institute of Technology (MIT), USA \cite{VArun2020}. The elements of RFocus can adjust the reflect signals toward specified direction and locations. Although the smart glass and RFocus have provided some design insights for the development of RIS, the research of RIS is still at an early stage \cite{XMu2021STAR}.

In academia, aiming to theoretically exploiting the benefits of RIS, there are much literature dedicates to the optimization problems in the RIS-assisted networks. For instance, in the RIS-assisted MEC networks, during a given mission period, the total completed task-input bits is maximized in \cite{MSun2021, ZChu2021, XHu2021}. From the simulation results, it can be observed that the computational performance of the MEC system is greatly improved with the aid of RIS. In order to minimize the energy consumption of all users in the RIS-assisted MEC, the phase shift, size of transmission data, transmission rate, power, transmission time and the decoding order are jointly optimized in \cite{ZLi2021}. Besides, the latency minimization and energy efficiency maximization problems in the RIS-assisted MEC are investigated in \cite{TBai2020} and \cite{CHuang2019}, respectively.

In the RIS-assisted UAV networks, the RIS can be deployed for overcoming the blockages and enhancing the achievable rate. To this end, in \cite{YPan2021}, the sum rate of users in the downlink is maximized by jointly optimizing UAV's trajectory, the phase shift of RIS, the allocation of THz sub-bands, and the power control. As expected, the largest sum rate is achieved by the proposed joint optimization algorithm. In addition, in order to guarantee the secure communication between the UAV and the legitimate user, the average secrecy rate is maximized by an iterative algorithm in \cite{ XPang2021}. Similarly, the secure transmission problem is investigated in the UAV and RIS assisted mmWave networks, where the near-optimal positions of RIS and UAV are obtained by an exhaustive searching method \cite{GSun2021}. Besides, in order to overcome the highly dynamic stochastic environments and reduce the computational complexity, the machine learning (ML) approaches have been utilized in the RIS-assisted UAV networks. For instance, in \cite{XLiu2021}, a decaying deep Q-network (D-DQN) based algorithm is proposed to minimize the energy consumption of the UAV by jointly optimizing the phase shift of RIS, UAV trajectory, decoding order, and power allocation. Simulation results show that the proposed D-DQN algorithm can strike a balance between accelerating training speed and converging to the local optimal, as well as avoiding oscillation. Furthermore, with the aim to maximize the energy efficiency of the UAV, the joint trajectory-phase-shift problems are tackled by the Double DQN (DDQN) and Deep Deterministic Policy Gradient (DDPG) algorithms in \cite{HMei2022}. It can be seen that the energy efficiency of the UAV is able to be greatly improved with the aid of RIS.

\section{System Model and Problem Formulation}
\subsection{System Model}

\begin{figure}[t]

\centering
\includegraphics[width =3.5 in]{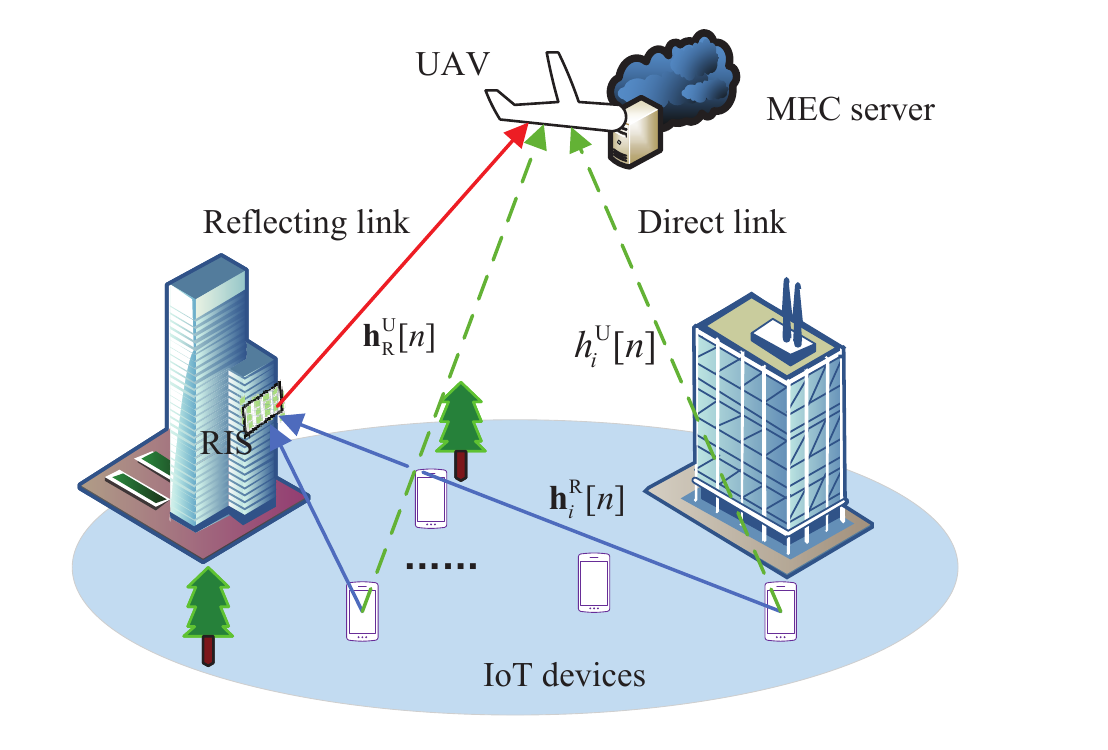}

\caption{The RIS-assisted UAV-enabled MEC system.}
\label{3}
\end{figure}

An RIS-assisted UAV-enabled MEC system is shown in Fig. 1, where an UAV equipped with an MEC server provides computing services to $I$ IoT devices. An RIS with $M$ reflection elements is installed on the surrounding building wall to assist IoT devices' task offloading. To ease of exposition, the IoT devices and reflection elements are denoted by \(i \in {\cal I} \buildrel \Delta \over = \left\{ {1,2,...,I} \right\}\) and \(m \in {\cal M} \buildrel \Delta \over = \left\{ {1,2,...,M} \right\}\), respectively. 

%
%
%
%
%
%
%
%
%
%
%
%

We suppose the system operates during an appointed mission period of $T$, which is divided into $N$ time slots and indexed by \(n \in {\cal N} \buildrel \Delta \over = \left\{ {1,2,...,N} \right\}\). The time slot length $t = T/N$ is sufficiently small so that the UAV flies through a small distance and the channel gain is approximately unchanged within each time slot. Without loss of generality, a 3D Cartesian coordinate system is adopted to describe the positions of UAV, RIS, and the IoT devices. Specifically, the horizontal position of the UAV at time slot $n$ can be represented as ${\bf{q}}[n] = ({x_{\rm{U}}}[n],{y_{\rm{U}}}[n])$. Similar to \cite{JJi2021,SJeong2018}, it is assumed that the UAV flies at a fixed altitude $H > 0$. The horizontal position and the altitude of the first element on the RIS \cite{HMei2021} are given by \({{\bf{w}}_{\rm{R}}} = \left( {{x_{\rm{R}}},{y_{\rm{R}}}} \right)\) and ${h_{\rm{R}}}$, respectively. In addition, the $i$-th IoT device is fixed at the ground with zero altitude and the horizontal position \({{\bf{w}}_i} = \left( {{x_i},{y_i}} \right)\) is known to the RIS and the UAV.

\subsubsection{Communication Model}
Since the UAV flies at a high altitude and the RIS is placed on the façade of a building, the communication link between the UAV and RIS is assumed to be a LoS channel. Thus, the channel gain between the UAV and the RIS at the $n$-th time slot can be given by
 \begin{equation}
{\bf{h}}_{\rm{R}}^{\rm{U}}[n] = \sqrt {\rho d_{{\rm{RU}}}^{ - 2}\left[ n \right]} \left[ {1,...,{e^{ - j\frac{{2\pi }}{\lambda }\left( {M - 1} \right)d{\varphi _{{\rm{RU}}}}[n]}}} \right],
 \end{equation}
where $\rho $ is the path loss at the reference ${D_0} = 1$m; ${d_{{\rm{RU}}}}[n] = \sqrt {{{(H - {h_{\rm{R}}})}^2} + {{\left\| {{\bf{q}}[n] - {{\bf{w}}_{\rm{R}}}} \right\|}^2}} $ denotes the distance between the UAV and the RIS at the $n$-th time slot; $d$ is the antenna separation;  $\lambda $ is the carrier wavelength; ${\varphi _{{\rm{RU}}}}[n] = {{\left( {{x_{\rm{R}}} - {x_{\rm{U}}}[n]} \right)} \mathord{\left/
 {\vphantom {{\left( {{x_{\rm{R}}} - {x_{\rm{U}}}[n]} \right)} {{d_{{\rm{RU}}}}[n]}}} \right.
 \kern-\nulldelimiterspace} {{d_{{\rm{RU}}}}[n]}}$ represents the cosine of the angle of departure (AoD) of the signal from the RIS to the UAV at the $n$-th time slot.

The direct links from the IoT devices to the UAV are assumed to be blocked by obstacles \cite{HMei2021, ARanjha2021,GLinghui2020}. Thus, the channel gain from the $i$-th IoT device to the UAV at the $n$-th time slot can be expressed as
 \begin{equation}
h_i^{\rm{U}}[n] = \sqrt {\rho d_{i{\rm{U}}}^{ - \varepsilon }[n]} {g_{i{\rm{U}}}},
 \end{equation}
where ${d_{i{\rm{U}}}}[n] = \sqrt {{{\left\| {{\bf{q}}[n] - {{\bf{w}}_i}} \right\|}^2} + {H^2}} $ is the distance between the UAV and the $i$-th IoT device at the $n$-th time slot; $\varepsilon $ is the path loss exponent and ${g_{i{\rm{U}}}}$ represents the random scattering component modeled by a zero-mean and unit-variance circularly symmetric complex Gaussian random variable.

For the communication links from the IoT devices to the RIS, we assume that they are Rician fading channels \cite{HMei2021}, which consist of the LoS and non-LoS (NLoS) components. Hence, the channel gain between the $i$-th IoT device and the RIS at the $n$-th time slot can be given by
 \begin{equation}
{\bf{h}}_i^{\rm{R}}\left[ n \right] = \sqrt {\rho d_{i{\rm{R}}}^{ - \gamma }\left[ n \right]} \left( {\sqrt {\frac{\beta }{{1 + \beta }}} {\bf{h}}_{i{\rm{R}}}^{{\rm{LoS}}} + \sqrt {\frac{1}{{1 + \beta }}} {\bf{h}}_{i{\rm{R}}}^{{\rm{NLoS}}}} \right),
 \end{equation}
where ${d_{i{\rm{R}}}} = \sqrt {{{\left\| {{{\bf{w}}_i} - {{\bf{w}}_{\rm{R}}}} \right\|}^2} + {h_{\rm{R}}}^2} $ is the distance between the $i$-th IoT device and the RIS;  $\gamma $ denotes the path loss exponent; $\beta $ represents the Rician factor; \({{\bf{h}}_{i{\rm{R}}}^{{\rm{LoS}}}}\) and \({{\bf{h}}_{i{\rm{R}}}^{{\rm{NLoS}}}}\) are the LoS component and NLoS component, respectively. For \({{\bf{h}}_{i{\rm{R}}}^{{\rm{LoS}}}}\), we have
 \begin{equation}
{\bf{h}}_{i{\rm{R}}}^{{\rm{LoS}}}\left[ n \right] = {\left[ {1,{e^{ - j\frac{{2\pi }}{\lambda }d{\varphi _{i{\rm{R}}}}}},...,{e^{ - j\frac{{2\pi }}{\lambda }(M - 1)d{\varphi _{i{\rm{R}}}}}}} \right]^T},
 \end{equation}
where \({\varphi _{i{\rm{R}}}} = {{\left( {{x_i} - {x_{\rm{R}}}} \right)} \mathord{\left/
 {\vphantom {{\left( {{x_i} - {x_{\rm{R}}}} \right)} {{d_{i{\rm{R}}}}}}} \right.
 \kern-\nulldelimiterspace} {{d_{i{\rm{R}}}}}}\) is the cosine of the angle of arrival (AoA) of the signal from the $i$-th IoT device to the RIS. The NLoS component \({{\bf{h}}_{i{\rm{R}}}^{{\rm{NLoS}}}}\) is the complex Gaussian distributed variable with zero mean and unit variance.

With the help of the RIS, it is possible to achieve the virtual LoS connections between the UAV and the IoT devices by adjusting the phase shift of RIS. Since the phase shift of each reflecting element of the RIS can be dynamically adjusted by a controller, in this paper, the phase-shift matrix of the RIS is modeled as
 \begin{equation}
{\bf{\Phi }}[n] = {\rm{diag}}\left\{ {{e^{j{\theta _1}[n]}},...,{e^{j{\theta _M}[n]}}} \right\},
 \end{equation}
where ${\theta _m}[n] \in \left[ {0,2\pi } \right]$ is the phase shift of the $m$-th RIS element at the  $n$-th time slot. Thus, the combined channel gain from the $i$-th IoT device to the UAV at the $n$-th time slot can be given by
 \begin{equation}
{h_i}[n] = h_i^{\rm{U}}[n] + {({\bf{h}}_i^{\rm{R}}[n])^H}{\bf{\Phi }}[n]{\bf{h}}_{\rm{R}}^{\rm{U}}[n].
 \end{equation}

Denote the bandwidth of the system as $B$. Benefiting from the partial offloading paradigm in MEC, at each time slot, the IoT devices can offload parts of their task-input data to the UAV with the aid of RIS. When the IoT devices offload tasks, the NOMA protocol is adopted to further improve the energy efficiency. To be specific, at each time slot, the IoT devices are ranked by the UAV in the ascending order of channel gain. Therefore, the order of the IoT devices for the UAV at the $n$-th time slot is denoted by \(\Pi  = \left\{ {{\pi _1}\left[ n \right],{\pi _2}\left[ n \right],...,{\pi _I}\left[ n \right]} \right\}\), where \({\pi _i}\left[ n \right]\) is the index of the IoT device with the $i$-th smallest channel gain to the UAV during time slot $n$.

After receiving the IoT device’s signal, the successive interference cancellation (SIC) technique is adopted by the UAV to decode signals from multiple IoT devices. When the UAV decodes the signal from IoT device \({\pi _i}\left[ n \right]\), the received signals from IoT device \({\pi _1}\left[ n \right]\) to IoT device ${\pi _{i - 1}}\left[ n \right]$ are regarded as interference. Thus, the offloading data rate of IoT device \({\pi _i}\left[ n \right]\) at the $n$-th time slot can be expressed as \cite{XZhang2020}
 \begin{equation}
R_{{\pi _i}}^{{\rm{off}}}\left[ n \right] = B\log \left( {1 + \frac{{{p_{{\pi _i}}}[n]{{\left| {{h_{{\pi _i}}}[n]} \right|}^2}}}{{\sum\nolimits_{j = 1}^{i - 1} {{p_{{\pi _j}}}[n]{{\left| {{h_{{\pi _j}}}[n]} \right|}^2} + {\sigma ^2}} }}} \right),
 \end{equation}
where ${p_{{\pi _i}}}[n]$ is the transmit power of IoT device ${\pi _i}[n]$ when offloading tasks to the UAV at the $n$-th time slot, and \({\sigma ^2}\) is the noise power. If the time slot index can be shown clearly in the variables, the order index ${\pi _i}[n]$ in the subscript is reduced to ${\pi _i}$ for ease of exposition.

\subsubsection{Computation Model}
Denote the task of each IoT device as a positive tuple $\left\{ {{L_i},{C_i}} \right\}$, where ${{L_i}}$ is the minimal amount of task-input bits of IoT device $i$ in the mission period, and ${{C_i}}$ is the CPU cycles required for computing 1-bit of task-input data. At each time slot, the IoT devices can perform local computing and task offloading simultaneously. Denote \(l_i^{{\rm{loc}}}\left[ n \right]\) as the task bits computed locally at IoT device $i$ during time slot $n$. Considering the limitation of IoT device's computation capacity, we have
 \begin{equation}
\frac{{l_i^{{\rm{loc}}}[n]{C_i}}}{t} \le {F_i},\forall i \in {\cal I},n \in {\cal N},
 \end{equation}
where \({F_i}\) is IoT device $i$'s maximum CPU-cycle frequency.

Similarly, denote the maximum CPU-cycle frequency of the UAV as \({F_{{\rm{UAV}}}}\). We have
 \begin{equation}
\frac{{\sum\nolimits_{i = 1}^I {l_i^{{\rm{UAV}}}\left[ n \right]{C_i}} }}{t} \le {F_{{\rm{UAV}}}},\forall n \in {\cal N},
 \end{equation}
where ${l_i^{{\rm{UAV}}}\left[ n \right]}$ denotes the amount of task bits that is computed by the UAV for IoT device $i$ at time slot $n$. Since the UAV can only compute the task that has been offloaded and received, we have
 \begin{equation}
R_i^{{\rm{off}}}\left[ n \right]t \ge l_i^{{\rm{UAV}}}\left[ n \right],\forall i \in {\cal I},n \in {\cal N}.
 \end{equation}
In addition, to meet all IoT devices' minimum computation requirements, we have
 \begin{equation}
\sum\limits_{n = 1}^N {\left( {l_i^{{\rm{loc}}}\left[ n \right] + l_i^{{\rm{UAV}}}\left[ n \right]} \right) \ge {L_i}} ,\forall i \in {\cal I}.
 \end{equation}

\subsubsection{Energy consumption model} The energy consumption of the IoT devices consists of two parts, i.e., the energy consumption for task offloading and that for local computing. Firstly, at time slot $n$, the task offloading energy consumption of IoT device $i$ is given by
 \begin{equation}
E_i^{{\rm{off}}}[n] = {p_i}[n]t.
 \end{equation}

Then, based on \cite{XZhang2020}, the local computing energy consumed by IoT device $i$ at the $n$-th time slot can be modeled as
 \begin{equation}
E_i^{{\rm{com}}}\left[ n \right] = \frac{{{\kappa _{{\rm{IoT}}}}{{\left( {l_i^{{\rm{loc}}}\left[ n \right]} \right)}^3}}}{{{t^2}}},
 \end{equation}
where ${\kappa _{{\rm{IoT}}}}$ is the effective capacitance coefficient of the IoT device that depends on the processor’ s chip architecture.
Thus, the total energy consumption of all IoT devices at time slot $n$ can be expressed as
 \begin{equation}
{E_{{\rm{IoT}}}}[n] = \sum\limits_{i = 1}^I {\left( {E_i^{{\rm{com}}}\left[ n \right] + E_i^{{\rm{off}}}\left[ n \right]} \right)}.
 \end{equation}

The UAV mounted by an MEC server flying in the sky provides computing services for the IoT devices. With a similar model to the IoT device, the computing energy consumption of the UAV at time slot $n$ is given by
 \begin{equation}
E_{\rm{U}}^{{\rm{com}}}\left[ n \right] = \sum\limits_{i = 1}^I {\frac{{{\kappa _{{\rm{UAV}}}}{{\left( {l_i^{{\rm{UAV}}}\left[ n \right]} \right)}^3}}}{{{t^2}}}},
 \end{equation}
where ${\kappa _{{\rm{UAV}}}}$ is the UAV’s effective capacitance coefficient.

In this paper, the flying energy consumption of the UAV is also taken into account. We deploy the fixed-wing UAV in the proposed system as an example, and its flying energy consumption at time slot $n$ can be modeled as \cite{XHu2020}
 \begin{equation}
E_{\rm{U}}^{{\rm{fly}}}\left[ n \right] = t\left( {{\tau _1}{v^3}\left[ n \right] + \frac{{{\tau _2}}}{{v\left[ n \right]}}} \right),
 \end{equation}
where $v[n] = {{\left\| {{\bf{q}}[n] - {\bf{q}}[n - 1]} \right\|} \mathord{\left/
 {\vphantom {{\left\| {{\bf{q}}[n] - {\bf{q}}[n - 1]} \right\|} t}} \right.
 \kern-\nulldelimiterspace} t}$ represents the speed of the UAV at time slot $n$. ${\tau _1}$ and ${\tau _2}$ are two parameters related to the UAV’s weight, wing area, wing span efficiency, and air density, etc.

Hence, the energy consumption of the UAV at time slot $n$ can be represented as
 \begin{equation}
{E_{{\rm{UAV}}}}[n] = E_{\rm{U}}^{{\rm{fly}}}[n] + E_{\rm{U}}^{{\rm{com}}}[n].
 \end{equation}

\subsection{Problem Formulation}
In this paper, we aim to maximize the energy efficiency of the RIS-assisted UAV-enabled MEC system. At each time slot, the total amount of completed task bits are comprised of the offloading task bits and those computed locally at the IoT devices. Thus, the total amount of completed task bits at time slot $n$ can be given by \cite{MLi2020, JZhang2020EE, XZhang2019}
 \begin{equation}
L\left[ n \right] = \sum\limits_{i = 1}^I {\left( {l_i^{{\rm{loc}}}\left[ n \right] + R_i^{{\rm{off}}}\left[ n \right]t} \right)}.
 \end{equation}

Meanwhile, the total energy consumption includes all IoT devices’ energy consumption and the UAV’s energy consumption, which can be expressed as
 \begin{equation}
E[n] = {E_{{\rm{IoT}}}}[n] + {E_{{\rm{UAV}}}}[n].
 \end{equation}
The energy efficiency is defined as the ratio of the total amount of completed task bits over the total energy consumption in the mission period. Thus, the energy efficiency maximization problem for RIS-assisted UAV-enabled systems can be formulated as
 \begin{subequations}
\begin{align}
&\mathop {\max }\limits_{\bf{z}} \frac{{\sum\nolimits_{n = 1}^N {L\left[ n \right]} }}{{\sum\nolimits_{n = 1}^N {E\left[ n \right]} }}\\
{\rm{s}}.{\rm{t}}.&\sum\limits_{n = 1}^N {(l_i^{{\rm{loc}}}\left[ n \right] + l_i^{{\rm{UAV}}}\left[ n \right]) \ge {L_i}} ,\forall i \in {\cal I},\\
{\rm{    }}&\frac{{\sum\nolimits_{i = 1}^I {l_i^{{\rm{UAV}}}\left[ n \right]} {C_i}}}{t} \le {F_{{\rm{UAV}}}},\forall n \in {\cal N},\\
{\rm{   }}&\frac{{l_i^{{\rm{loc}}}[n]{C_i}}}{t} \le {F_i},\forall i \in {\cal I},n \in {\cal N},\\
{\rm{   }}&R_i^{{\rm{off}}}[n]t \ge l_i^{{\rm{UAV}}}[n],\forall i \in {\cal I},n \in {\cal N},\\
{\rm{   }}&\left| {{\theta _m}[n]} \right| = 1,\forall m \in {\cal M},n \in {\cal N},\\
{\rm{    }}&{\bf{q}}[1] = {{\bf{q}}_0},{\bf{q}}[N + 1] = {{\bf{q}}_F},\\
{\rm{   }}&||v[n]|| \le {V_{{\rm{Max}}}},\forall n \in {\cal N},
\end{align}
 \end{subequations}
where \({\bf{z}} = \left\{ {l_i^{{\rm{loc}}}[n],l_i^{{\rm{UAV}}}[n],{p_i}[n],{\theta _m}[n],{\bf{q}}[n]} \right\}\). Constraint (20b) ensures the minimum computation requirements of IoT devices can be satisfied. Constraints (20c) and (20d) mean that the workloads of UAV and IoT devices cannot exceed their maximum CPU frequencies. Constraint (20f) represents the feasible set of RIS's phase shift. Constraint (20g) is UAV's initial and ﬁnal horizontal locations. Constraint (20h) represents that the speed of UAV must be less than the maximum speed.

\section{Solution to the Formulated Problem}
Due to the fractional structure of the objective function, and the closely coupled optimization
variables in (20), it is difficult to obtain the globally optimal solution in polynomial time. To tackle these challenges, we propose an iterative algorithm with a double-loop structure to maximize the energy efficiency and optimize the bit allocation \(l_i^{{\rm{loc}}}[n]\) and \(l_i^{{\rm{UAV}}}[n]\), transmit power \({{p_i}[n]}\), phase shift of RIS \({{\theta _m}[n]}\), and the UAV trajectory \({{\bf{q}}[n]}\). In the outer loop, the Dinkelbach's method is exploited to handle the fraction programming and obtain the energy efficiency. With the given energy efficiency, the coupled variables are iteratively optimized by the block coordinate descent (BCD) method in the inner loop.

Firstly, we equivalently transform problem (20) into the following parametric problem:
 \begin{equation}
\begin{array}{*{20}{l}}
{\mathop {\max }\limits_{{\bf{z}},\alpha } \sum\limits_{n = 1}^N {L[n]}  - \alpha \sum\limits_{n = 1}^N {E[n]} ,}\\
{{\rm{s}}.{\rm{t}}.(20{\rm{b}}) - (20{\rm{h}}),}
\end{array}
 \end{equation}
where $\alpha $ is the introduced auxiliary parameter. Assuming \({\alpha ^*}\) is the optimal objective value of problem (20), we have the following Theorem.

\textbf{\emph{Theorem 1:}} The optimal solution \({{\bf{z}}^*}\) of problem (20) can be obtained if and only if
 \begin{equation}
\mathop {\max }\limits_{{\bf{z}}} \left( {\sum\limits_{n = 1}^N {L[n]}  - {\alpha ^*}\sum\limits_{n = 1}^N {E[n]} } \right) = 0.
 \end{equation}

\emph{Proof:} We prove the theorem by the sufficient and necessary conditions. On one hand, according to (22), we have $\left( {\sum\limits_{n = 1}^N {L\left[ n \right]\left( {{{\bf{z}}^*}} \right)}  \!-\! {\alpha ^*}\sum\limits_{n = 1}^N {E[n]\left( {{{\bf{z}}^*}} \right)} } \right)\! = \!0.$ For any other ${\bf{z}}$, $\left( {\sum\nolimits_{n = 1}^N {L[n]} \left( {\bf{z}} \right) - {\alpha ^*}\sum\nolimits_{n = 1}^N {E[n]\left( {\bf{z}} \right)} } \right) < 0$. Thus, $\frac{{\sum\nolimits_{n = 1}^N {L\left[ n \right]\left( {{{\bf{z}}^*}} \right)} }}{{\sum\nolimits_{n = 1}^N {E\left[ n \right]\left( {{{\bf{z}}^*}} \right)} }} > \frac{{\sum\nolimits_{n = 1}^N {L\left[ n \right]\left( {\bf{z}} \right)} }}{{\sum\nolimits_{n = 1}^N {E\left[ n \right]\left( {\bf{z}} \right)} }}$ and \({{\bf{z}}^*}\) is the optimal solution of the energy efficiency maximization problem (20).

On the other hand, if \({{\bf{z}}^*}\) is the optimal solution of (20), we have
 \begin{equation}
\mathop {\max }\limits_{{{\bf{z}}^*}} \frac{{\sum\nolimits_{n = 1}^N {L\left[ n \right]} }}{{\sum\nolimits_{n = 1}^N {E\left[ n \right]} }} = {\alpha ^*}.
  \end{equation}
Then, the equation (22) can be obtained from (23) after simple transformation. Theorem 1 is proved. $ \hfill{}\blacksquare $

However, the optimal \({\alpha ^*}\) cannot be obtained in advance. Hence we propose an iterative algorithm based on the Dinkelbach's method to update $\alpha $. The details can be seen in Algorithm 1.

\begin{algorithm}[t]
\caption{Dinkelbach's algorithm for maximizing the energy efficiency}
\centering
\begin{tabular}{p{8cm}}
\noindent\hangafter=1\setlength{\hangindent}{1.2em}1. Initialize ${\bf{z}}$, iterative number $k = 1$.

\noindent\hangafter=1\setlength{\hangindent}{1.2em}2. \textbf{repeat}:

\noindent\hangafter=1\setlength{\hangindent}{2.4em}3. \hspace{1em} Solve problem (21) for given ${\alpha ^{(k)}}$, and obtain the optimal solution \({{\bf{z}}^{(k)}}\).

\noindent\hangafter=1\setlength{\hangindent}{2.4em}4. \hspace{1em} Calculate $F({\alpha ^{(k)}}) = {\left| {\sum\limits_{n = 1}^N {L[n]}  - \alpha \sum\limits_{n = 1}^N {E[n]} } \right|^{(k)}}$.

\noindent\hangafter=1\setlength{\hangindent}{1.2em}5. \hspace{1em} \textbf{if} $F({\alpha ^{(k)}}) \le \delta $ then

\noindent\hangafter=1\setlength{\hangindent}{1.2em}6. \hspace{2em} ${\alpha ^*} = \frac{{\sum\nolimits_{n = 1}^N {L{{[n]}^{(k)}}} }}{{\sum\nolimits_{n = 1}^N {E{{[n]}^{(k)}}} }}$; ${{\bf{z}}^*} = {{\bf{z}}^{(k)}}$; break.

\noindent\hangafter=1\setlength{\hangindent}{1.2em}7. \hspace{1em} \textbf{else}
\({\alpha ^{(k + 1)}} = \frac{{\sum\nolimits_{n = 1}^N {L{{[n]}^{(k)}}} }}{{\sum\nolimits_{n = 1}^N {E{{[n]}^{(k)}}} }}\); $k = k + 1$.

\noindent\hangafter=1\setlength{\hangindent}{1.2em}8. \textbf{Until} $k \ge {N_{\max }}$.

\noindent\hangafter=1\setlength{\hangindent}{1.2em}9. \textbf{Output}: the optimal energy efficiency \({\alpha ^*}\) and the corresponding solution \({{\bf{z}}^*}\).
\end{tabular}
\end{algorithm}

In Algorithm 1, problem (21) needs to be solved with given \({\alpha ^{(k)}}\). However, with given energy efficiency \({\alpha ^{(k)}}\), problem (21) is still non-convex due to the coupling among UAV trajectory \({{\bf{q}}[n]}\), phase shift of RIS \({{\theta _m}[n]}\), and other optimization variables. Therefore, in order to tackle problem (21), it is decomposed into three subproblems by adopting the BCD technique, namely, bit allocation and transmit power optimization, phase shift optimization, and UAV trajectory optimization. And then an iterative algorithm is proposed to solve them in an alternating manner.

\subsection{Bit Allocation and Transmit Power Optimization}
With given phase shift \({{\theta _m}[n]}\) and UAV trajectory \({{\bf{q}}[n]}\), the bit allocation and transmit power optimization problem can be reformulated from (21) as
 \begin{subequations}
\begin{align}
&\mathop {\max }\limits_{l_i^{{\rm{loc}}}[n],l_i^{{\rm{UAV}}}[n],{p_i}[n]} \sum\limits_{n = 1}^N {L[n]}  - \alpha \sum\limits_{n = 1}^N {E[n]} \\
{\rm{s}}.{\rm{t}}.&(20{\rm{b}}) - (20{\rm{e}}).
\end{align}
\end{subequations}

\textbf{\emph{Theorem 2:}} Define \({S_{{\pi _i}}}[n] \!=\! \sum\nolimits_{j = 1}^i {{p_{{\pi _j}}}[n]{{\left| {{h_{{\pi _j}}}[n]} \right|}^2}\! +\! {\sigma ^2}} .\) The task offloading energy consumption of all IoT devices in the mission period can be expressed as
 \begin{equation}
\sum\limits_{n = 1}^N {\sum\limits_{i = 1}^I {{p_{{\pi _i}}}[n]t} } \! = \!\sum\limits_{n = 1}^N {\sum\limits_{i = 1}^I {t\left( {\frac{1}{{{h_{{\pi _i}}}{{[n]}^2}}} \!-\! \frac{1}{{{h_{{\pi _{i + 1}}}}{{[n]}^2}}}} \right)} } {2^{\frac{{\sum\limits_{j = 1}^i {R_{{\pi _j}}^{{\rm{off}}}[n} ]}}{B}}}.
 \end{equation}

\emph{Proof:} According to (7), it can be found that \({2^{R_{{\pi_i}}^{{\rm{off}}}[n]/B}} - 1 = \frac{{{p_{{\pi_i}}}[n]{{\left| {{h_{{\pi_i}}}[n]} \right|}^2}}}{{\sum\nolimits_{j = 1}^{i - 1} {{p_{{\pi_j}}}[n]{{\left| {{h_{{\pi_j}}}[n]} \right|}^2}}  + {\sigma ^2}}}\). Then, with the definition of ${S_{{\pi_i}}}[n]$, the recursion expression can be obtained as
 \begin{equation}
{S_{{\pi_{i - 1}}}}[n]\left( {{2^{R_{{\pi_i}}^{\rm{off}}[n]/B}} - 1} \right) = {S_{{\pi_i}}}[n] - {S_{{\pi_{i - 1}}}}[n].
 \end{equation}
Then we can further obtain \(\;{S_{{\pi_i}}}[n] = {S_{{\pi_{i - 1}}}}[n]{2^{R_{{\pi_i}}^{\rm{off}}[n]/B}},i \in {\cal I}\backslash \{ 1\} \). If letting ${S_{{\pi_1}}}[n] = {p_{{\pi_1}}}[n]{\left| {{h_{{\pi_1}}}[n]} \right|^2} + {\sigma ^2}$, we have \({S_{{\pi_i}}}[n] = {2^{\frac{{\sum\nolimits_{j = 1}^i {R_{{\pi_j}}^{{\rm{off}}}[n]} }}{B}}},i \in I\backslash \{ 1\} \). Thus, the transmit power of IoT device \({\pi_i}[n]\) at time slot $n$ can be obtained as
 \begin{equation}
{p_{{\pi_i}}}[n] = \frac{{{S_{{\pi_i}}}[n] - {S_{{\pi_{i - 1}}}}[n]}}{{{{\left| {{h_{{\pi_i}}}[n]} \right|}^2}}},i \in {\cal I}\backslash \{ 1\}.
 \end{equation}
Equation (27) is also valid for $i = 1$ if we set \(\sum\nolimits_{j = 1}^0 {R_{{\pi_j}}^{{\rm{off}}}[n]}  = 0\). Therefore, the total offloading energy consumption of IoT devices during the mission period can be expressed as (25). Theorem 2 is proved. $ \hfill{}\blacksquare $

By substituting (25) into the objective function of problem (24) and defining \({{\bf{\Omega }}_1} = \left\{ {l_i^{{\rm{loc}}}\left[ n \right],l_i^{{\rm{UAV}}}\left[ n \right],R_{{r_i}}^{{\rm{off}}}\left[ n \right]} \right\}\), problem (24) can be transformed into
 \begin{subequations}
\begin{align}
&\mathop {\max }\limits_{{{\bf{\Omega }}_1}} \sum\limits_{n = 1}^N {\sum\limits_{i = 1}^I {\left( {l_i^{{\rm{loc}}}[n] + R_{{\pi _i}}^{{\rm{off}}}[n]t - \alpha \frac{{{\kappa _{{\rm{IoT}}}}{{\left( {l_i^{{\rm{loc}}}[n]} \right)}^3}}}{{{t^2}}}} \right.} } \notag \\
&\left. { \!-\! \alpha t(\frac{1}{{{h_{{\pi _i}}}{{[n]}^2}}}\! -\! \frac{1}{{{h_{{\pi _{i + 1}}}}{{[n]}^2}}}){2^{\frac{{\sum\nolimits_{j = 1}^i {R_{{\pi _j}}^{{\rm{off}}}[n]} }}{B}}}\! -\! \alpha \frac{{{\kappa _{{\rm{UAV}}}}{{\left( {l_i^{{\rm{UAV}}}[n]} \right)}^3}}}{{{t^2}}}} \right)\\
{\rm{s}}.{\rm{t}}.&(20{\rm{b}}) - (20{\rm{d}}),\\
&R_{{\pi _i}}^{{\rm{off}}}\left[ n \right]t \ge l_{{\pi _i}}^{{\rm{UAV}}}\left[ n \right].
\end{align}
\end{subequations}

Remark 1: It can be found that the transmit power optimization \({{p_i}\left[ n \right]}\) in (24) is transformed to the optimization of offloading rate \({R_{{\pi _i}}^{{\rm{off}}}\left[ n \right]}\). Constraint (20e) is also expressed as (28c) related to \({R_{{\pi _i}}^{{\rm{off}}}\left[ n \right]}\). By solving problem (28), the offloading rate can be obtained, and then the transmit power of IoT device ${\pi_i}\left[ n \right]$ can be calculated according to (27). Define a function $\Psi \left( {i,n} \right)$ to indicate the decoding order of IoT device $i$ for the UAV in time slot $n$ when adopting the NOMA protocol. Thus, the transmit power of IoT device $i$ at time slot $n$ can be obtained as ${p_{{\pi_{\Psi \left( {i,n} \right)}}}}[n]$.

Note that the objective function of (28) is convex with respect to $l_i^{{\rm{loc}}}[n],l_i^{{\rm{UAV}}}[n]$, and $R_{{\pi_i}}^{\rm{off}}[n]$. Meanwhile, the constraints of problem (28) are linear. Thus, problem (28) is a convex optimization problem, which can be solved by CVX. In order to obtain more insights, we further derive its closed-form solution in the following Theorem 3.

\textbf{\emph{Theorem 3:}} For problem (28), the optimal bit allocation and transmit power can be expressed as
 \begin{equation}
l_i^{{\rm{loc}}}{[n]^*} = \sqrt {\left( {1 + {\omega _i} + \frac{{{\psi _{i,n}}C}}{t}} \right)\frac{{{t^2}}}{{3\alpha {\kappa _{{\rm{IoT}}}}}}} ,
 \end{equation}
 \begin{equation}
l_i^{{\rm{UAV}}}{[n]^*} = \sqrt {\left( {{\omega _i} + {\xi _{{r_{\Psi \left( {i,n} \right)}},n}} + \frac{{{\zeta _n}C}}{t}} \right)\frac{{{t^2}}}{{3\alpha {\kappa _{{\rm{UAV}}}}}}} ,
 \end{equation}
 \begin{equation}
{p_{{\pi _i}}}{[n]^*} \!=\! \frac{{B{\xi _{i,n}}t}}{{\alpha \ln 2}}\left( {\frac{{{h_{{\pi _{i + 1}}}}{{[n]}^2}}}{{{h_{{\pi _{i + 1}}}}{{[n]}^2}\! -\! {h_{{\pi _i}}}{{[n]}^2}}}\! -\! \frac{{{h_{{\pi _{i - 1}}}}{{[n]}^2}}}{{{h_{{\pi _i}}}{{[n]}^2} \!- \!{h_{{\pi _{i - 1}}}}{{[n]}^2}}}} \right),
 \end{equation}
where \({\left\{ {{\omega _i}} \right\}_{i \in {\cal I}}},{\left\{ {{\psi _{i,n}}} \right\}_{i \in {\cal I},n \in {\cal N}}},{\left\{ {{\varsigma _n}} \right\}_{n \in {\cal N}}},{\left\{ {{\xi _{i,n}}} \right\}_{i \in {\cal I},n \in {\cal N}}}\) are Lagrange multipliers associated with constraints (20b)-(20d) and (28c), respectively.

\emph{Proof:} The Lagrangian function of problem (28) can be expressed as
 \begin{equation}
\begin{array}{*{20}{l}}
{L({{\bf{\Omega }}_2}) = \sum\limits_{n = 1}^N {\sum\limits_{i = 1}^I {\left( {l_i^{{\rm{loc}}}[n] + tR_{{\pi _i}}^{{\rm{off}}}[n] - \alpha \frac{{{\kappa _{{\rm{IoT}}}}{{\left( {l_i^{{\rm{loc}}}[n]} \right)}^3}}}{{{t^2}}}} \right.} } }\\
{\left. { - \alpha t\left( {\frac{1}{{{h_{{\pi _i}}}{{[n]}^2}}} - \frac{1}{{{h_{{\pi _{i + 1}}}}{{[n]}^2}}}} \right){2^{\frac{{\sum\nolimits_{j = 1}^i {R_{{\pi _j}}^{{\rm{off}}}[n]} }}{B}}} - \alpha \frac{{{\kappa _{{\rm{UAV}}}}{{\left( {l_i^{{\rm{UAV}}}[n]} \right)}^3}}}{{{t^2}}}} \right)}\\
\begin{array}{l}
 + \sum\limits_{i = 1}^I {{\omega _i}\left( {\sum\limits_{n = 1}^N {\left( {l_i^{{\rm{loc}}}[n] + l_i^{{\rm{UAV}}}[n]} \right)}  - {L_i}} \right)} \\
 + \sum\limits_{n = 1}^N {\sum\limits_{i = 1}^I {{\psi _{i,n}}\left( {\frac{{l_i^{{\rm{loc}}}[n]C}}{t} - {F_i}} \right)} }
\end{array}\\
\begin{array}{l}
 + \sum\limits_{n = 1}^N {{\varsigma _n}\left( {\frac{{\sum\limits_{i = 1}^I {l_i^{{\rm{UAV}}}[n]} {C_i}}}{t} - {F_{{\rm{UAV}}}}} \right)} \\
 + \sum\limits_{n = 1}^N {\sum\limits_{i = 1}^I {{\xi _{i,n}}\left( {l_{{\pi _i}}^{{\rm{UAV}}}[n] - R_{{\pi _i}}^{{\rm{off}}}[n]t} \right)} } ,
\end{array}
\end{array}
 \end{equation}
where ${{\bf{\Omega }}_2} = \left\{ {l_i^{{\rm{loc}}}[n],l_i^{{\rm{UAV}}}[n],R_{{\pi_i}}^{{\rm{off}}}[n],{\omega _i},{\psi _{i,n}},{\varsigma _n},{\xi _{i,n}}} \right\}$. Then, the derivations of \(L({{\bf{\Omega }}_2})\) with respect to ${l_i^{{\rm{loc}}}[n],l_i^{{\rm{UAV}}}[n]}$ and $R_{{\pi_i}}^{{\rm{off}}}[n]$ can be given by
 \begin{equation}
\frac{{\partial L\left( {{{\bf{\Omega }}_2}} \right)}}{{\partial l_i^{{\rm{loc}}}[n]}} = 1 - \frac{{3\alpha {\kappa _{{\rm{IoT}}}}{{\left( {l_i^{{\rm{loc}}}[n]} \right)}^2}}}{{{t^2}}} + {\omega _i} + \frac{{{\psi _{i,n}}C}}{t},
 \end{equation}
  \begin{equation}
\frac{{\partial L\left( {{{\bf{\Omega }}_2}} \right)}}{{\partial l_i^{{\rm{UAV}}}[n]}} =  - \frac{{3\alpha {\kappa _{{\rm{UAV}}}}{{\left( {l_i^{{\rm{UAV}}}[n]} \right)}^2}}}{{{t^2}}} + {\omega _i} + {\xi _{{\pi _{\Psi \left( {i,n} \right)}},n}} + \frac{{{\zeta _n}C}}{t},
 \end{equation}
   \begin{equation}
\begin{array}{*{20}{l}}
\begin{array}{l}
\frac{{\partial L\left( {{{\bf{\Omega }}_2}} \right)}}{{\partial R_{{\pi _i}}^{{\rm{off}}}[n]}} =  - \frac{{\alpha t\ln 2}}{B}\left( {\frac{1}{{{h_{{\pi _i}}}{{[n]}^2}}} - \frac{1}{{{h_{{\pi _{i + 1}}}}{{[n]}^2}}}} \right){2^{\frac{{\sum\nolimits_{j = 1}^i {R_{{\pi _j}}^{{\rm{off}}}[n} ]}}{B}}}\\
 \;\; \;\; \;\; \;\; \;\; \;\; \;\; \;\; \;\; \;\; + t - {\xi _{i,n}}t.
\end{array}
\end{array}
 \end{equation}

According to the KKT conditions, by setting the derivations of \(L({{\bf{\Omega }}_2})\) to zero, the optimal solution to problem (28) can be obtained as (29), (30) and
   \begin{equation}
{2^{\frac{{\sum\nolimits_{j = 1}^i {R_{{\pi_j}}^{{\rm{off}}}[n]} }}{B}}} = \frac{{1 - {\xi _{i,n}}}}{{\frac{{\alpha \ln 2}}{B}\left( {\frac{1}{{{h_{{\pi_i}}}{{[n]}^2}}} - \frac{1}{{{h_{{\pi_{i + 1}}}}{{[n]}^2}}}} \right)}}.
 \end{equation}
And then, by substituting (36) into (27), the optimal transmit power of IoT device ${\pi_i}[n]$ at time slot $n$ can be obtained as (31). Theorem 3 is proved.  $ \hfill{}\blacksquare $

\subsection{Phase Shift Optimization}
For given bit allocation \(l_i^{{\rm{loc}}}[n]\) and \(l_i^{{\rm{UAV}}}[n]\), transmit power \({{p_{{\pi _i}}}[n]}\), and UAV trajectory \({{\bf{q}}\left[ n \right]}\), problem (21) can be reformulated as
 \begin{subequations}
\begin{align}
&{\mathop {\max }\limits_{{\theta _m}[n]} \sum\limits_{n = 1}^N {\sum\limits_{i = 1}^I {Bt\log \left( {1 + \frac{{{p_{{\pi_i}}}[n]{{\left| {{h_{{\pi_i}}}[n]} \right|}^2}}}{{\sum\nolimits_{j = 1}^{i - 1} {{p_{{\pi_j}}}[n]{{\left| {{h_{{\pi_j}}}[n]} \right|}^2} + {\sigma ^2}} }}} \right)} } },\\
&{{\rm{s}}.{\rm{t}}.\left| {{\theta _m}[n]} \right| = 1,\forall m \in {\cal M},n \in {\cal N}.}
\end{align}
\end{subequations}

It can be observed that problem (37) is actually an offloading rate maximization problem. When the bit allocation, transmit power, and UAV trajectory are given, the RIS only has impacts on the offloading rate by controlling the phase shift of reflected signals. Moreover, due to the objective function, problem (37) is still non-convex and difficult to tackle directly.

To solve problem (37), we first transform it into a more tractable form. Define ${\bf{h}}_{{\pi_i}}^{{\rm{RIS}}}[n] = {\bf{h}}_{{\pi_i}}^{\rm{R}}{[n]^H}{\rm{diag}}\left( {{\bf{h}}_{\rm{R}}^{\rm{U}}[n]} \right)$. The channel gain between IoT device ${\pi_i}[n]$ and the UAV at time slot $n$ can be expressed as
   \begin{equation}
\begin{array}{*{20}{l}}
\begin{array}{l}
{\left| {{h_{{\pi _i}}}[n]} \right|^2} = {\left| {h_{{\pi _i}}^{\rm{U}}[n] + {{({\bf{h}}_{{\pi _i}}^{\rm{R}}[n])}^H}{\rm{diag}}({\bf{\Phi }}[n]){\bf{h}}_{\rm{R}}^{\rm{U}}[n]} \right|^2}\\
\;\;\;\;\;\;\;\;\;\;\;\;\;\; = {\rm{Tr}}\left( {{{\bf{H}}_{{\pi _i}}}[{\rm{n}}]{\bf{\Theta }}[n]} \right){\rm{ + }}{\left( {h_{{\pi _i}}^{\rm{U}}[n]} \right)^2},
\end{array}
\end{array}
 \end{equation}
where \({{\bf{H}}_{{\pi_i}}}[n] = \left( {\begin{array}{*{20}{c}}
{{\bf{h}}_{{\pi_i}}^{{\rm{RIS}}}{{[n]}^H}{\bf{h}}_{{\pi_i}}^{{\rm{RIS}}}[n]}&{{\bf{h}}_{{\pi_i}}^{{\rm{RIS}}}{{[n]}^H}{\bf{h}}_{{\pi_i}}^{\rm{U}}[n]}\\
{{\bf{h}}_{{\pi_i}}^{\rm{U}}[n]{\bf{h}}_{{\pi_i}}^{{\rm{RIS}}}[n]}&0
\end{array}} \right)\), and \({\bf{\Theta }}[n] = {\bf{\bar \Phi }}[n]{\left( {{\bf{\bar \Phi }}[n]} \right)^H}\) is a positive semidefinite matrix with \({\bf{\bar \Phi }}[n] = {[{e^{j{\theta _1}[n]}},...,{e^{j{\theta _M}[n]}},x]^T}\). $x$ is an auxiliary scalar. By substituting (38) into the objective function of problem (37), we have
   \begin{equation}
\begin{array}{*{20}{l}}
\begin{array}{l}
\log (1 + \frac{{{p_{{\pi _i}}}[n]{{\left| {{h_{{\pi _i}}}[n]} \right|}^2}}}{{\sum\nolimits_{j = 1}^{i - 1} {{p_{{\pi _j}}}[n]{{\left| {{h_{{\pi _j}}}[n]} \right|}^2} + {\sigma ^2}} }})\\
 = {\log _2}\left( {\sum\limits_{j = 1}^i {{p_{{\pi _j}}}[n]} \left( {{\rm{Tr}}({{\bf{H}}_{{\pi _j}}}[{\rm{n}}]{\bf{\Theta }}[n]){\rm{ + }}{{\left( {h_{{\pi _j}}^{\rm{U}}[n]} \right)}^2}} \right) + {\sigma ^2}} \right)
\end{array}\\
\begin{array}{l}
 - {\log _2}\left( {\sum\limits_{j = 1}^{i - 1} {{p_{{\pi _j}}}[n]} \left( {{\rm{Tr}}({{\bf{H}}_{{\pi _j}}}[{\rm{n}}]{\bf{\Theta }}[n]){\rm{ + }}{{\left( {h_{{\pi _j}}^{\rm{U}}[n]} \right)}^2}} \right) + {\sigma ^2}} \right)\\
 = W_1^i[n] - W_2^i[n].
\end{array}
\end{array}
 \end{equation}

Thus, problem (37) can be transformed into
 \begin{subequations}
\begin{align}
&{\mathop {\max }\limits_{{\bf{\Theta }}[n]} \sum\limits_{n = 1}^N {\sum\limits_{i = 1}^I {\left( {W_1^i[n] - W_2^i[n]} \right)} } }\\
&{{\rm{s}}.{\rm{t}}.{\Theta _{m,m}}[n] = 1,\forall m \in {\cal M},n \in {\cal N},}\\
\;\;\;\; &{ {{\rm{rank}}({\bf{\Theta }}[n]) = 1,\forall n \in {\cal N}.}}
\end{align}
\end{subequations}

In problem (40), the objective function is still non-convex with respect to the phase-shift-related variables ${{\bf{\Theta }}[n]}$. Moreover, the rank one constraint (40c) makes the problem more difficult to solve. To tackle these challenges, it can be found that the objective function of (40) is the difference of concave functions, which can be handled by exploiting the DC programming technique \cite{XHu2021}. Thus, in the $\left( {l + 1} \right)$-th iteration of the DC programming, the second term of the objective function is approximated by its linear upper bound, i.e.,
   \begin{equation}
\begin{array}{*{20}{l}}
\begin{array}{l}
W_2^i[n] \!\le\! \frac{{\sum\nolimits_{j = 1}^{i - 1} {{p_{{\pi _j}}}\left[ n \right]} \left\langle {\left( {{\bf{\Theta }}[n] \!-\! {\bf{\Theta }}{{[n]}^{(l)}}} \right),{{\left. {{\nabla _{\bf{\Theta }}}{\rm{Tr}}\left( {{{\bf{H}}_{{\pi _j}}}[n]{\bf{\Theta }}[n]} \right)} \right|}_{{\bf{\Theta }} = {{\bf{\Theta }}^{(l)}}}}} \right\rangle }}{{\ln 2\left( {\sum\nolimits_{j = 1}^{i - 1} {{p_{{\pi _j}}}} \left[ n \right]\left( {{\rm{Tr}}\left( {{{\bf{H}}_{{\pi _j}}}[n]{\bf{\Theta }}{{[n]}^{(l)}}} \right) + {{\left| {h_{{\pi _j}}^D[n]} \right|}^2}} \right) + {\sigma ^2}} \right)}}\\
 \;\;\;\;\;\;\;\;\;\;\;\;\;\;+ {(W_2^i[n])^{(l)}} = \tilde W_2^i[n]
\end{array}
\end{array}
 \end{equation}

As for the rank one constraint (40c), we introduce the semi-definite programming relaxation (SDR) technique by dropping the rank-one constraint \cite{QWu2019}. To this end, problem (40) can be expressed as
 \begin{subequations}
\begin{align}
&\mathop {\max }\limits_{{\bf{\Theta }}[n]} \sum\limits_{n = 1}^N {\sum\limits_{i = 1}^I {\left( {W_1^i[n] - \tilde W_2^i[n]} \right)} } \\
{\rm{s}}.{\rm{t}}.&{{\bf{\Theta }}_{m,m}}[n] = 1,\forall m \in {\cal M},n \in {\cal N},\\
&{\bf{\Theta }}[n] \succ 0,\forall n \in {\cal N},
\end{align}
 \end{subequations}

Note that  problem (42) is a standard convex semi-definite programming (SDP) and can be solved via classic convex toolboxes, such as the SDP solver in the CVX tool \cite{QWu2019}. Then, we iteratively update \({\bf{\Theta }}[n]\) by solving problem (42) to tighten the lower bound of the objective function of (40) until convergence. During the iteration, when the rank of \({\bf{\Theta }}[n]\) is larger than one, the Gaussian randomization method is adopted to recover ${\bf{\Phi }}[n]$ from \({\bf{\Theta }}[n]\) \cite{SHuang2021}. To be specific, a set of vectors which obey the distribution of \({\cal C}{\cal N}\left( {0,{\bf{\Theta }}[n]} \right)\) is first generated. And then, we choose the candidate which maximizes the objective function of (42) as the phase shift of RIS. The detailed optimization procedure of RIS's phase shift is outlined in Algorithm 2.

\begin{algorithm}[t]
\caption{Proposed algorithm for solving problem (37)}
\centering
\begin{tabular}{p{8cm}}
\noindent\hangafter=1\setlength{\hangindent}{1.2em}1. Initialize $\left\{ {{\theta _m}[n]} \right\}$, iterative number $l = 1$.

\noindent\hangafter=1\setlength{\hangindent}{1.2em}2. \textbf{while} \(\left| {\sum\limits_{n = 1}^N {\sum\limits_{i = 1}^I {{{\left( {W_1^i[n] - W_2^i[n]} \right)}^{(l + 1)}}} } } \right.\)

\noindent\hangafter=1\setlength{\hangindent}{1.2em} \;\; \;\; \;\; \;\; \;\; \;\;\(\left. { - \sum\limits_{n = 1}^N {\sum\limits_{i = 1}^I {{{\left( {W_1^i[n] - W_2^i[n]} \right)}^{(l)}}} } } \right| \ge \delta \) \textbf{do}

\noindent\hangafter=1\setlength{\hangindent}{2.4em}3. \hspace{1em} Calculate \(\tilde W_2^i[n]\) based on (41).

\noindent\hangafter=1\setlength{\hangindent}{2.4em}4. \hspace{1em} Solve the convex problem (42) to obtain \({{\bf{\Theta }}[n]}\).

\noindent\hangafter=1\setlength{\hangindent}{2.4em}5. \hspace{1em} Utilize the Gaussian randomization technique to recover \({{\theta _m}[n]}\).

\noindent\hangafter=1\setlength{\hangindent}{1.2em}6. \hspace{1em} $l \leftarrow l + 1$

\noindent\hangafter=1\setlength{\hangindent}{1.2em}7. \textbf{end while}
\end{tabular}
\end{algorithm}

\subsection{UAV Trajectory Optimization}

Finally, supposing the bit allocation \({l_i^{{\rm{loc}}}[n]}\) and \({l_i^{{\rm{UAV}}}[n]}\), transmit power \({{p_{{\pi _i}}}[n]}\), and phase shift \({{\theta _m}[n]}\) are given, the UAV trajectory optimization problem can be reformulated as
   \begin{subequations}
\begin{align}
\mathop {\max }\limits_{{\bf{q}}[n]} &\sum\limits_{n = 1}^N {\sum\limits_{i = 1}^I {Bt\log \left( {1 + \frac{{{p_{{\pi_i}}}[n]{{\left| {{h_{{\pi_i}}}[n]} \right|}^2}}}{{\sum\nolimits_{j = 1}^{i - 1} {{p_{{\pi_j}}}[n]{{\left| {{h_{{\pi_j}}}[n]} \right|}^2} + {\sigma ^2}} }}} \right)} } \notag \\
&- t\alpha \sum\limits_{n = 1}^N {\left( {{\tau _1}{v^3}[n] + \frac{{{\tau _2}}}{{v[n]}}} \right)} \\
{\rm{s}}{\rm{.t}}{\rm{.}}&R_{{\pi_i}}^{\rm{off}}[n]t \ge l_{{\pi_i}}^{\rm{UAV}}[n],i \in {\cal I},n \in {\cal N},\\
&{\rm{(20g)}},{\rm{(20h)}}.
\end{align}
 \end{subequations}

Due to constraint (43b) and the objective function, problem (43) is non-convex and hard to solve. Thus, we introduce several auxiliary variables and then leverage the SCA technique to transform problem (43) into a convex problem. Firstly, define \(M_1^i[n] = B\log \left( {\sum\nolimits_{j = 1}^i {{p_{{\pi_j}}}[n]{{\left| {{h_{{\pi_j}}}[n]} \right|}^2}}  + {\sigma ^2}} \right)\) and \(M_2^i[n] = B\log \left( {\sum\nolimits_{j = 1}^{i - 1} {{p_{{\pi_j}}}[n]{{\left| {{h_{{\pi_j}}}[n]} \right|}^2}}  + {\sigma ^2}} \right)\). The offloading rate of IoT device \({\pi_i}[n]\) at time slot $n$ can be given by $R_{{\pi_i}}^{\rm{off}}[n] = M_1^i[n] - M_2^i[n]$. With optimal phase shift \({{\theta _m}[n]}\), the channel gain between IoT device \({\pi_i}[n]\) and the UAV at time slot $n$ can be expressed as \cite{SLi2020}
   \begin{equation}
\begin{array}{*{20}{l}}
\begin{array}{l}
{h_{{\pi _i}}}[n] = h_{{\pi _i}}^{\rm{U}}[n] + {({\bf{h}}_{{\pi _i}}^{\rm{R}}[n])^H}{\bf{\Phi }}[n]{\bf{h}}_{\rm{R}}^{\rm{U}}[n]\\
\;\;\;\;\;\;\;\;\;\; = \frac{{\sqrt \rho  \left| {{g_{{\pi _i}{\rm{U}}}}} \right|}}{{d_{{\pi _i}{\rm{U}}}^{\varepsilon /2}[n]}} + \frac{{\sqrt \rho  \sum\nolimits_{m = 1}^M {\left| {{h_{{\pi _i}{\rm{R}},m}}} \right|} }}{{{d_{{\rm{RU}}}}[n]}}.
\end{array}
\end{array}
 \end{equation}
Then, the auxiliary variables ${u_{{\pi_i}}}[n]$ and $w[n]$ are introduced with \({d_{{\pi_i}{\rm{U}}}}[n] \le {u_{{\pi_i}}}[n],{d_{{\rm{RU}}}}[n] \le w[n]\). By replacing the term \({d_{{\pi_i}{\rm{U}}}}[n]\) and \({d_{{\rm{RU}}}}[n]\) in $M_1^i[n]$ with ${u_{{\pi_i}}}[n]$ and $w[n]$, we can obtain
  \begin{equation}
\tilde M_1^i[n] = B\log \left( {\sum\limits_{j = 1}^i {{p_{{\pi_i}}}\left[ n \right]\Xi {{\left( {{u_{{\pi_i}}}[n],w[n]} \right)}^2} + {\sigma ^2}} } \right),
\end{equation}
where \(\Xi \left( {{u_{{\pi_i}}}[n],w[n]} \right) = \frac{{\sqrt \rho  \left| {{g_{{\pi_i}{\rm{U}}}}} \right|}}{{u_{{\pi_i}{\rm{U}}}^{\varepsilon /2}[n]}} + \frac{{\sqrt \rho  \sum\nolimits_{m = 1}^M {\left| {{h_{{\pi_i}{\rm{R}},m}}} \right|} }}{{w[n]}}\). Similarly, for $M_2^i[n]$, we have
  \begin{equation}
\tilde M_2^i[n] = B\log \left( {\sum\limits_{j = 1}^{i - 1} {{p_{{\pi_i}}}\left[ n \right]\Xi {{\left( {{u_{{\pi_i}}}[n],w[n]} \right)}^2} + {\sigma ^2}} } \right).
\end{equation}

In addition, for the term $v[n]$ in the denominator of objective function, another auxiliary variable $\bar v[n]$ is introduced with $\bar v[n] \le v[n]$. Thus, problem (43) can be transformed into
 \begin{subequations}
\begin{align}
\mathop {\max }\limits_{{u_{{\pi_i}}}[n],w[n],\bar v[n],{\bf{q}}[n]} &\sum\limits_{n = 1}^N {\sum\limits_{i = 1}^I {t\left( {\tilde M_1^i[n] - \tilde M_2^i[n]} \right)} } \notag \\
 &- t\alpha \sum\limits_{n = 1}^N {\left( {{\tau _1}{v^3}[n] + \frac{{{\tau _2}}}{{\bar v[n]}}} \right)} \\
{\rm{s}}{\rm{.t}}{\rm{.}}&{d_{{\pi_i}{\rm{U}}}}[n] \le {u_{{\pi_i}}}[n],\\
&{d_{{\rm{RU}}}}[n] \le w[n],\\
&\bar v[n] \le v[n],\\
&\left( {\tilde M_1^i[n] - \tilde M_2^i[n]} \right)t \ge l_{{\pi_i}}^{\rm{UAV}}[n],\\
&{\bf{q}}[1] = {{\bf{q}}_0},{\bf{q}}[N + 1] = {{\bf{q}}_F},\\
&||v[n]|| \le {V_{\rm{Max}}}.
\end{align}
 \end{subequations}

\textbf{\emph{Theorem 4:}} \({\tilde M_1^i[n]}\) is convex with respect to \({u_{{\pi_i}}}[n]\) and \(w[n]\).


\emph{Proof:} At first, we define the function \(f(x,y) = \log \left( {\frac{{{k_1}}}{{{x^\varepsilon }}} + \frac{{{k_2}}}{{{x^{\varepsilon /2}}y}} + \sum\nolimits_{m = 1}^i {{{\left( {{{{b_m}} \mathord{\left/
 {\vphantom {{{b_m}} y}} \right.
 \kern-\nulldelimiterspace} y} + {c_m}} \right)}^2}} \! + \!{k_3}} \right)\), with \({c_m} \!\ge\! 0\), and  \({k_1},{k_2},{k_3},{b_m}> 0\). Then, the second-order partial derivatives of $f(x,y)$ can be given by
 \begin{equation}
\frac{{{\partial ^2}f}}{{\partial {x^2}}} = \frac{1}{{A\ln 2}}\left( {\frac{{\varepsilon \left( {\varepsilon  + 1} \right){k_1}}}{{{x^{\varepsilon  + 2}}}} + \frac{{{k_2}\varepsilon \left( {\varepsilon /2 + 1} \right)}}{{2{x^{\varepsilon /2 + 1}}y}}} \right),
 \end{equation}
 \begin{equation}
\frac{{{\partial ^2}f}}{{\partial {y^2}}} = \frac{1}{{A\ln 2}}\left( {\frac{{2{k_2}}}{{{x^{\varepsilon /2}}{y^3}}} + \frac{{4b_m^2}}{{{y^3}}} + \frac{{2{b_m}{c_m}}}{{{y^2}}}} \right),
 \end{equation}
 \begin{equation}
\frac{{{\partial ^2}f}}{{\partial x\partial y}} = \frac{1}{{2A\ln 2}}\frac{\varepsilon }{{{x^{\varepsilon /2 + 1}}}}\frac{{{k_2}}}{{{y^2}}},
 \end{equation}
where \(A = \frac{{{k_1}}}{{{x^\varepsilon }}} + \frac{{{k_2}}}{{{x^{\varepsilon /2}}y}} + \sum\nolimits_{m = 1}^i {{{\left( {{{{b_m}} \mathord{\left/
 {\vphantom {{{b_m}} y}} \right.
 \kern-\nulldelimiterspace} y} + {c_m}} \right)}^2}}  + {k_3}\). Then, we have $\frac{{{\partial ^2}f}}{{\partial {x^2}}} > 0$ and \(\frac{{{\partial ^2}f}}{{\partial {x^2}}}\frac{{{\partial ^2}f}}{{\partial {y^2}}} - \frac{{{\partial ^2}f}}{{\partial x\partial y}}\frac{{{\partial ^2}f}}{{\partial y\partial x}} > 0\). Thus, the Hessian matrix of $f(x,y)$ is positive definite and $f(x,y)$ is convex with respect to $x$ and $y$. Therefore, \({\tilde M_1^i[n]}\) is convex with respect to \({u_{{\pi_i}}}[n]\) and \(w[n]\).  $ \hfill{}\blacksquare $

Since any convex function is globally lower bounded by its first-order Taylor expansion at any point, based on Theorem 4, a lower-bound of \({\tilde M_1^i[n]}\) at the $l$-th iteration of SCA can be expressed as
 \begin{equation}
\begin{array}{*{20}{l}}
\begin{array}{l}
\tilde M_1^i[n] \ge \hat M_1^i[n] = \log {A_i}[n] + \frac{{{B_i}[n]}}{{{A_i}[n]\ln 2}}({u_{{\pi _i}}}[n] - {u_{{\pi _i}}}{[n]^{(l)}})\\
 \;\;\;\;\;\;\;\;\;+ \frac{{{C_i}[n]}}{{{A_i}[n]\ln 2}}(w[n] - w{[n]^l}),
\end{array}
\end{array}
 \end{equation}
where
 \begin{equation}
{A_i}[n] = \sum\nolimits_{j = 1}^i {{p_{{\pi _j}}}[n]\Xi {{\left( {{u_{{\pi _j}}}{{[n]}^{(l)}},w{{[n]}^{(l)}}} \right)}^2}}  + {\sigma ^2},
 \end{equation}
  \begin{equation}
{B_i}[n] =  - {p_{{\pi _i}}}[n]\Xi \left( {{u_{{\pi _i}}}{{[n]}^{(l)}},w{{[n]}^{(l)}}} \right)\frac{{\varepsilon \sqrt \rho  \left| {{g_{{\pi _i}U}}} \right|}}{{u_{{\pi _i}}^{\varepsilon /2 + 1}[n]}},
 \end{equation}
  \begin{equation}
{C_i}[n] =  - {p_{{\pi _i}}}[n]\Xi \left( {{u_{{\pi _i}}}{{[n]}^{(l)}},w{{[n]}^{(l)}}} \right)\frac{{\sqrt \rho  \sum\nolimits_{m = 1}^M {\left| {{h_{{\pi _i}R,m}}} \right|} }}{{{w^2}[n]}}.
 \end{equation}

Similarly, constraints (47b)-(47d) can be approximated as
 \begin{equation}
{({d_{{\pi_i}U}}[n])^2} + {({u_{{\pi_i}}}{[n]^{(l)}})^2} - 2{u_{{\pi_i}}}{[n]^{(l)}}{u_{{\pi_i}}}[n] \le 0,
 \end{equation}
 \begin{equation}
{({d_{RU}}[n])^2} + {(w{[n]^{(l)}})^2} - 2w{[n]^{(l)}}w[n] \le 0,
 \end{equation}
  \begin{equation}
\begin{array}{*{20}{l}}
\begin{array}{l}
\bar v{[n]^2}{t^2} + {\left\| {{\bf{q}}{{[n]}^{(l)}} \!-\! {\bf{q}}{{[n - 1]}^{(l)}}} \right\|^2}\\
 \!-\! 2{\left( {{\bf{q}}{{[n]}^{(l)}} - {\bf{q}}{{[n - 1]}^{(l)}}} \right)^T}\left( {{\bf{q}}[n] \!-\! {\bf{q}}[n - 1]} \right) \!\le\! 0.
\end{array}
\end{array}
 \end{equation}

After the above operations, problem (47) can be reformulated as
 \begin{subequations}
\begin{align}
\mathop {\max }\limits_{{u_{{\pi _i}}}[n],w[n],\bar v[n],{\bf{q}}[n]} &\sum\limits_{n = 1}^N {\sum\limits_{i = 1}^I {\left( {\hat M_1^i[n] - \tilde M_2^i[n]} \right)} }  \notag \\
&- t\alpha \sum\limits_{n = 1}^N {\left( {{\tau _1}{v^3}[n] + \frac{{{\tau _2}}}{{\bar v[n]}}} \right)} \\
{\rm{s}}.{\rm{t}}.&\left( {\hat M_1^i[n] - \tilde M_2^i[n]} \right)t \ge l_{{\pi _i}}^{{\rm{UAV}}}[n],\\
&{\bf{q}}[1] = {{\bf{q}}_0},{\bf{q}}[N + 1] = {{\bf{q}}_F},\\
&|v[n]|| \le {V_{{\rm{Max}}}},\\
&(55) - (57).
\end{align}
 \end{subequations}

Problem (58) is a convex optimization problem, and thus can be solved efficiently by standard solvers. Then, based to the SCA method, the auxiliary variables \({{u_{{\pi _i}}}[n],w[n],\bar v[n]}\) and UAV trajectory \({{\bf{q}}[n]}\) are iteratively updated to tighten the lower bound of the objective function of (43) until convergence. Finally, problem (43) is effectively solved and the optimized UAV trajectory can be obtained. Denote the objective function of (43) at the $l$-th iteration as $E{\left( {{\bf{q}}[n]} \right)^{(l)}}$. The proposed algorithm for UAV trajectory optimization is outlined in Algorithm 3.

\begin{algorithm}[t]
\caption{Proposed algorithm for solving problem (43)}
\centering
\begin{tabular}{p{8cm}}
\noindent\hangafter=1\setlength{\hangindent}{1.2em}1. Initialize the vector \({u_{{\pi _i}}}{[n]^{(0)}},w{[n]^{(0)}},\bar v{[n]^{(0)}},{\bf{q}}{[n]^{(0)}}\) and set the iteration number as \(l=0\).

\noindent\hangafter=1\setlength{\hangindent}{1.2em}2. \textbf{while} $\left| {E{{({\bf{q}}[n])}^{(l + 1)}} - E{{({\bf{q}}[n])}^{(l)}}} \right| \ge \delta $ \textbf{do}

\noindent\hangafter=1\setlength{\hangindent}{2.4em}3. \hspace{1em} Calculate $\hat M_1^i[n]$ and $\tilde M_2^i[n]$ based on (51) and (46), respectively.

\noindent\hangafter=1\setlength{\hangindent}{2.4em}4. \hspace{1em} Solve the convex optimization problem (58) and obtain \({u_{{\pi _i}}}{[n]^{(l)}},w{[n]^{(l)}},\bar v{[n]^{(l)}},{\bf{q}}{[n]^{(l)}}\).

\noindent\hangafter=1\setlength{\hangindent}{1.2em}5. \hspace{1em} \(l \leftarrow l + 1\)

\noindent\hangafter=1\setlength{\hangindent}{1.2em}6. \textbf{end while}
\end{tabular}
\end{algorithm}

\subsection{Joint Optimization of Bit allocation, Transmit Power, Phase Shift, and UAV Trajectory }
Based on the obtained solutions to the three subproblems, the proposed BCD algorithm for solving problem (21) with given energy efficiency is summarized in Algorithm 4. Therefore, according to Algorithm 1, the original non-convex problem (20) can be effectively solved by iteratively updating the energy efficiency in the outer-loop and jointly optimizing the bit allocation, transmit power, phase shift, and UAV trajectory in the inner-loop via Algorithm 4.

\begin{algorithm}[t]
\caption{BCD algorithm for solving Problem (21)}
\centering
\begin{tabular}{p{8cm}}
\noindent\hangafter=1\setlength{\hangindent}{1.2em} 1. Initialize ${\alpha _k}$, iterative number $l = 1$.

\noindent\hangafter=1\setlength{\hangindent}{1.2em} 2. \textbf{repeat}:

\noindent\hangafter=1\setlength{\hangindent}{2.4em} 3. \hspace{1em}Solve the convex problem (24) to obtain \(l_i^{{\rm{loc}}}{\left[ n \right]^{(l)}},l_i^{{\rm{UAV}}}{\left[ n \right]^{(l)}}\), and \({p_i}{\left[ n \right]^{(l)}}\) for given \({{\theta _m}[n]}\) and \({{\bf{q}}[n]}\).

\noindent\hangafter=1\setlength{\hangindent}{2.4em} 4. \hspace{1em}Exploit Algorithm 2 to obtain \({\theta _m}{[n]^{(l)}}\) according to the updated \(l_i^{{\rm{loc}}}{\left[ n \right]^{(l)}},l_i^{{\rm{UAV}}}{\left[ n \right]^{(l)}}\), \({p_i}{\left[ n \right]^{(l)}}\), and the given \({{\bf{q}}[n]}\).

\noindent\hangafter=1\setlength{\hangindent}{2.4em} 5.\hspace{1em} Run Algorithm 3 to obtain \({\bf{q}}{[n]^{(l)}}\) with updated \(l_i^{{\rm{loc}}}{\left[ n \right]^{(l)}},l_i^{{\rm{UAV}}}{\left[ n \right]^{(l)}}\), \({p_i}{\left[ n \right]^{(l)}}\), and \({\theta _m}{[n]^{(l)}}\).

\noindent\hangafter=1\setlength{\hangindent}{1.2em} 6.\hspace{1em} Calculate \(F\left( {{{\bf{z}}^{(l)}}} \right) \!=\! \left| {\sum\nolimits_{n \!=\! 1}^N {L[n} ] \!-\! {\alpha _k}\sum\nolimits_{n \!= \!1}^N {E[n]} } \right|\).

\noindent\hangafter=1\setlength{\hangindent}{1.2em} 7.\hspace{1em} Update the iterative index $l = l + 1$.

\noindent\hangafter=1\setlength{\hangindent}{1.2em} 8. \textbf{Until}: $l > {N_{\max }}$ or \(\left| {F({{\bf{z}}^{(l + 1)}}) - F({{\bf{z}}^{(l)}})} \right| \le \delta {\rm{ }}\).

\noindent\hangafter=1\setlength{\hangindent}{1.2em} 9. \textbf{Output}: bit allocation \(l_i^{{\rm{loc}}}{[n]^*}\) and \(l_i^{{\rm{UAV}}}{[n]^*}\), transmit power \({p_i}{[n]^*}\), phase shift \({\theta _m}{[n]^*}\), and UAV trajectory \({\bf{q}}{[n]^*}\).
\end{tabular}
\end{algorithm}

\subsection{Convergence and Complexity Analysis}
According to Algorithm 1 and Algorithm 4, the energy efficiency maximization problem can be solved via an iterative algorithm with a double-loop structure. In the outer loop, the energy efficiency $\alpha $ is updated via the Dinkelbach's method. The convergence of the Dinkelbach's method has been proved in \cite{WDinkelbach}. The update rule of $\alpha $ can be given by
\begin{equation}
\begin{array}{*{20}{l}}
\begin{array}{l}
{\alpha ^{(k + 1)}} = \frac{{\sum\nolimits_{n = 1}^N {L{{[n]}^{(k)}}} }}{{\sum\nolimits_{n = 1}^N {E{{[n]}^{(k)}}} }}\\
 \;\;\;\;\;\;\;\;\;\;= {\alpha ^{(k)}} - \frac{{\sum\nolimits_{n = 1}^N {L{{[n]}^{(k)}} - {\alpha ^{(k)}}\sum\nolimits_{n = 1}^N {E{{[n]}^{(k)}}} } }}{{ - \sum\nolimits_{n = 1}^N {E{{[n]}^{(k)}}} }}\\
\;\;\;\;\;\;\;\;\;\; = {\alpha ^{(k)}} - \frac{{F({\alpha ^{(k)}})}}{{F'({\alpha ^{(k)}})}},
\end{array}
\end{array}
\end{equation}
which implies the super-linear convergence rate of Dinkelbach's method \cite{ZAlessio20171}. The inner loop is the joint optimization of bit allocation, transmit power, phase shift, and UAV trajectory. A BCD-based algorithm is proposed to solve problem (21) and the convergence is verified in Theorem 5.

\textbf{\emph{Theorem 5:}} Algorithm 4 monotonically increases the objective function of problem (21) at each iteration and finally converges.

\emph{Proof:} Given the energy efficiency ${\alpha _k}$, the objective function of problem (21) is denoted as $F({{\bf{z}}^{(l)}})$ after the $l$-th iteration of Algorithm 3. Then, we have the inequalities shown at the top of this page.

\begin{figure*}
    \begin{equation}
 \begin{array}{l}
F\left( {{{\bf{z}}^{(l)}}} \right) \le F\left( {l_i^{{\rm{loc}}}{{[n]}^{(l + 1)}},l_i^{{\rm{UAV}}}{{[n]}^{(l + 1)}},{p_i}{{[n]}^{(l + 1)}},{\theta _m}{{[n]}^{(l)}},{\bf{q}}{{[n]}^{(l)}}} \right)\\
\;\;\;\;\;\;\;\;\;\;\;\;\; \le F\left( {l_i^{{\rm{loc}}}{{[n]}^{(l + 1)}},l_i^{{\rm{UAV}}}{{[n]}^{(l + 1)}},{p_i}{{[n]}^{(l + 1)}},{\theta _m}{{[n]}^{(l + 1)}},{\bf{q}}{{[n]}^{(l)}}} \right)\\
\;\;\;\;\;\;\;\;\;\;\;\;\; \le F\left( {l_i^{{\rm{loc}}}{{[n]}^{(l + 1)}},l_i^{{\rm{UAV}}}{{[n]}^{(l + 1)}},{p_i}{{[n]}^{(l + 1)}},{\theta _m}{{[n]}^{(l + 1)}},{\bf{q}}{{[n]}^{(l + 1)}}} \right).
\end{array}
    \end{equation}
    \hrulefill
\end{figure*}

The first inequality holds because \(l_i^{{\rm{loc}}}{[n]^{(l + 1)}},l_i^{{\rm{UAV}}}{[n]^{(l + 1)}}\), and \({p_i}{[n]^{(l + 1)}}\) is the optimal solution to problem (21) when given \({{\theta _m}{{[n]}^{(l)}}}\) and \({{\bf{q}}{{[n]}^{(l)}}}\). The second inequality follows the fact that \({{\theta _m}{{[n]}^{(l + 1)}}}\) is solved via Algorithm 2 with given \(l_i^{{\rm{loc}}}{[n]^{(l + 1)}},l_i^{{\rm{UAV}}}{[n]^{(l + 1)}}\), \({p_i}{[n]^{(l + 1)}}\), and \({\bf{q}}{[n]^{(l)}}\). The third inequality holds because the solution to problem (43) does not decrease the objective function of problem (21). Since \(F(\bf{z})\) is upper bounded, Algorithm 4 must converge after limited numbers of iterations.  $ \hfill{}\blacksquare $

Therefore, with Theorem 5 and the update rule of $\alpha $, the proposed energy efficiency maximization algorithm converges within a limited number of iterations \cite{ZAlessio2017}.

The computational complexity of Algorithm 1 depends on the number of iterations required for convergence, and the complexity required to solve problem (21) via Algorithm 4 in the inner-loop.
Thus, we first analysis the computational complexity of Algorithm 4 which consists of three subproblems. For subproblem 1, the interior point method can be applied since it is a standard convex optimization problem. Hence, the computational complexity of the $\tilde \varepsilon $-optimal solution can be expressed as ${\cal O}\left( {\ln (1/\tilde \varepsilon ){n^3}} \right)$, where $n = 3IN$ is the number of decision variables. The subproblem 2 is solved by DC programming and SDP. Denote the number of iterations as ${L_1}$. The complexity of Algorithm 2 can be expressed as \({{\cal O}_2} \buildrel \Delta \over = {\cal O}\left( {{L_1}\ln (1/\tilde \varepsilon ){{\left( {N\left( {M + 1} \right)} \right)}^{3.5}}} \right)\) \cite{XMu2021STAR}. For subproblem 3, the SCA technique is applied to optimize the UAV trajectory. The computation complexity can be given by ${\cal O}\left( {{L_2}\ln (1/\tilde \varepsilon )n_3^3} \right)$, where ${L_2}$ is the iteration number and  ${n_3} = KN$. Supposing the iteration number of outer loop is ${L_3}$, the computational complexity of the overall algorithm can be expressed as \({\cal O}\left( {{L_3}\left( {\ln (1/\tilde \varepsilon ){n^3} + {{\cal O}_2} + {L_2}\ln (1/\tilde \varepsilon )n_3^3} \right)} \right)\). It can be observed that the proposed energy efficiency maximization algorithm is in polynomial complexity.

\section{Simulation Results}
In this section, we present simulation results and compare the proposed energy efficiency maximization algorithm with other baselines. In the simulations, we consider an RIS-assisted UAV-enabled MEC system, where the UAV flies from ${q_0} = [30,50]$m to ${q_F} = [30,0]$m with the maximum speed of \({V_{{\rm{Max}}}} = 10\)m/s providing computing services for $I = 6$ IoT devices. The horizontal position of the first element on the RIS is $[50,25]$m, and the height is 20m. We assume the IoT devices have the same amount of task-input bits, i.e., ${L_1} = {L_2} = ... = {L_I}$. Some other simulation parameters are listed in Table I.

\begin{table}[t]\small
\caption{Simulation Parameters \cite{XQin2021}\cite{XHu2021}}
\centering

\begin{tabular}{|c|c|c|c|}
\hline
\bfseries Parameters & \bfseries Values & \bfseries Parameters & \bfseries Values \\
\hline
Total bandwidth &30 MHz & ${\tau _1},{\tau _2}$ & 0.00614, 15.976 \\
\hline
\({V_{{\rm{Max}}}}\) & 10 m/s & \({\beta _0}\) & -30 dB \\
\hline
Number of time slot & 20 & Noise power & -50 dBm \\
\hline
$H$ & 40 m & ${h_R}$ & 20 m \\
\hline
 ${F_i}$ & 3 GHz  & ${F_{\rm{UAV}}}$ & 12 GHz\\
\hline
$\gamma $ & 2.8 & $\varepsilon $ &  3.5\\
\hline
\end{tabular}

\end{table}

\begin{figure}[t]

\centering
\includegraphics[width =3.5in]{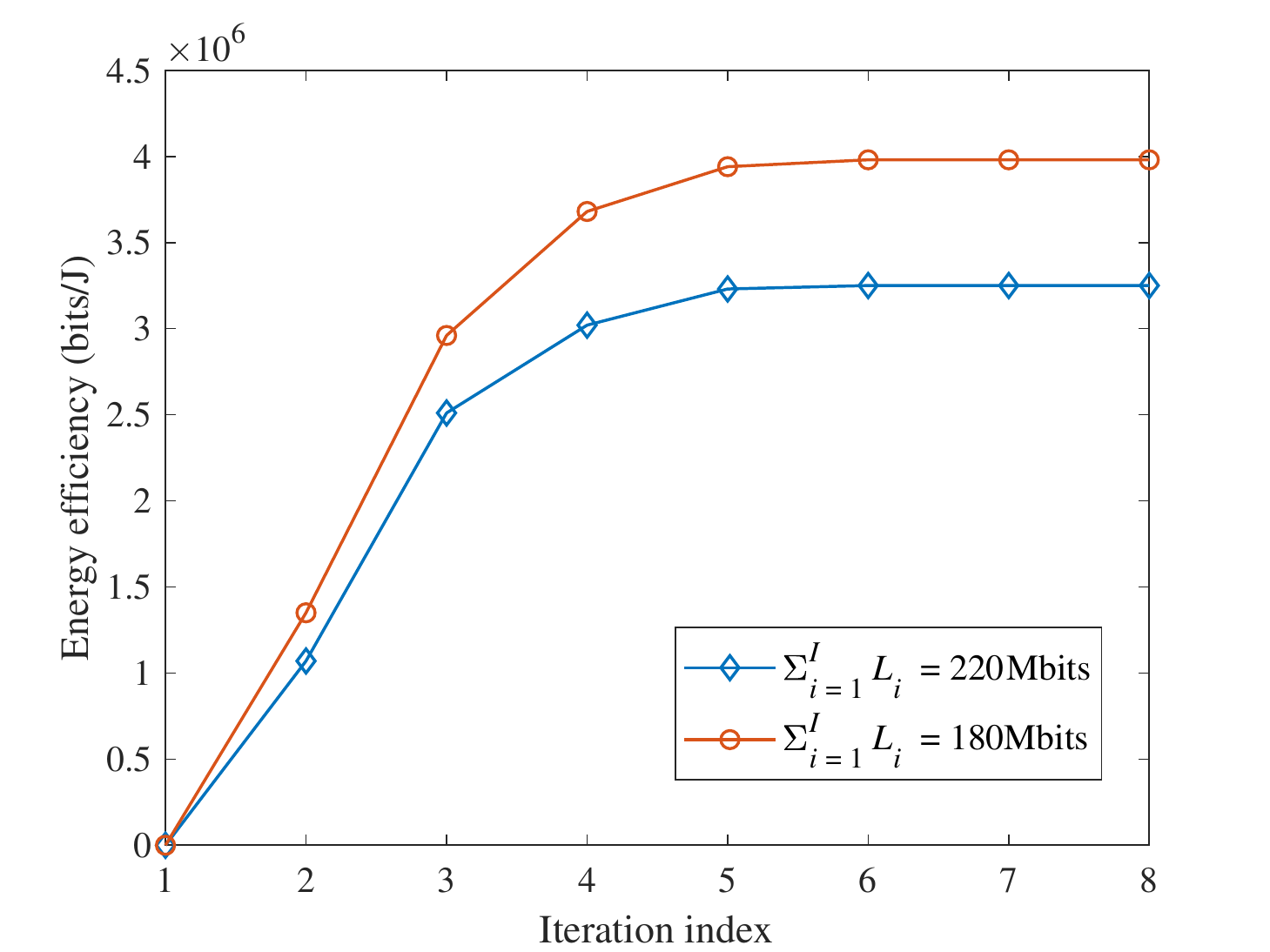}

\caption{Energy efficiency versus iteration index under two different size of IoT devices' total task-input bits.}
\label{3}

\end{figure}


Fig. 2 demonstrates the convergence evolution of the proposed energy efficiency maximization algorithm in the RIS-assisted UAV-enabled MEC system with $T = 10$ sec. It can be seen that the energy efficiency is increased rapidly at first and converges after around 5-6 iterations. Moreover, under different sizes of IoT devices' total task-input bits, the proposed algorithm still converges fast.

\begin{figure}[t]

\centering
\includegraphics[width =3.5in]{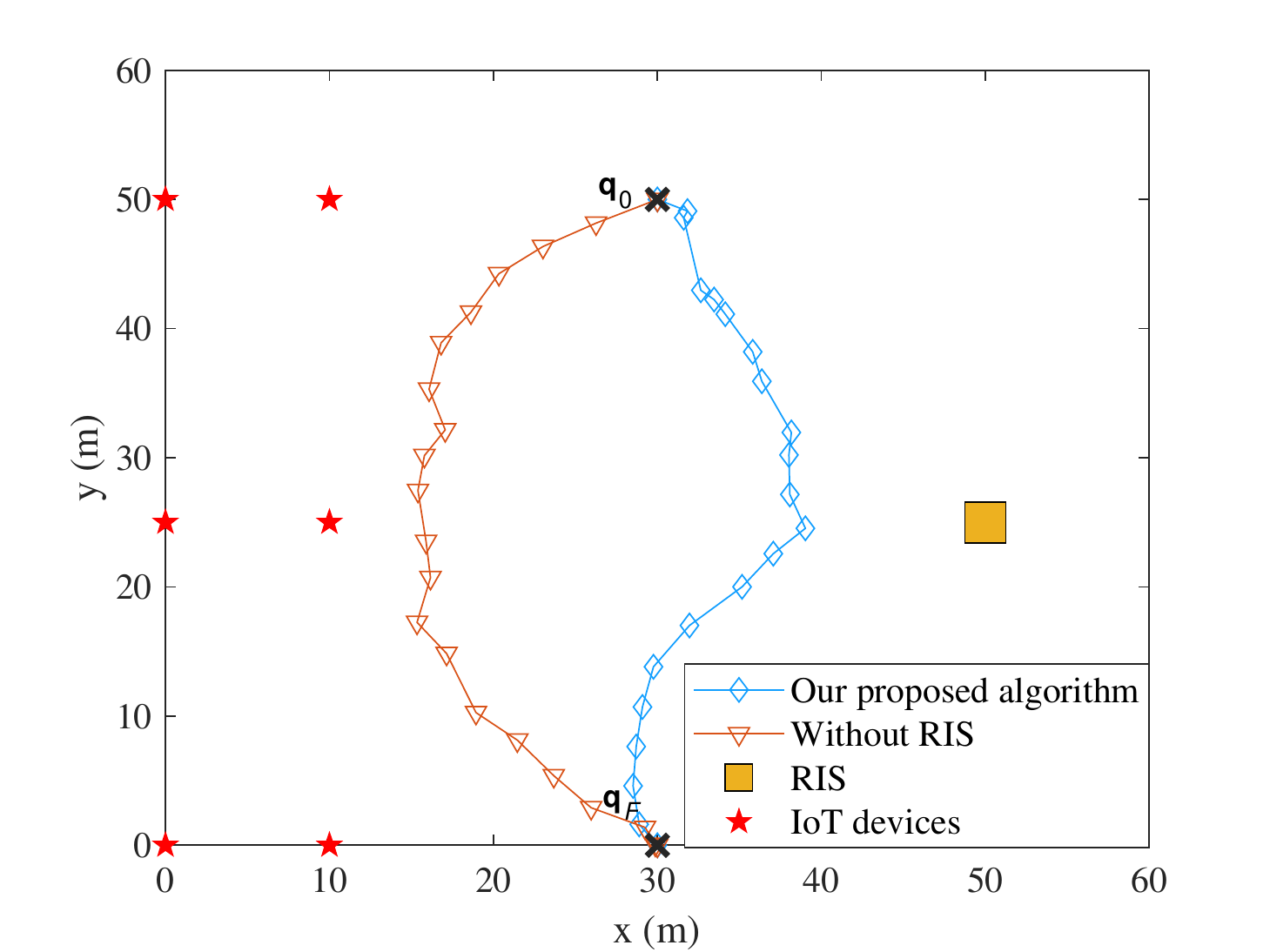}

\caption{UAV trajectories under two different scenarios: our proposed algorithm and the scheme without RIS.}
\label{3}
\vspace{-0.1in}
\end{figure}

Fig. 3 illustrates the trajectory of UAV under two different scenarios, i.e., our proposed algorithm and the scheme without RIS. Under the scheme without RIS, it can be observed that the UAV tends to fly closer to the IoT devices in order to achieve higher channel gains. On the contrary, in our proposed RIS-assisted UAV-enabled MEC system, we observe that the UAV tends to fly closer to the RIS. The reasons behind this can be explained as follows. When the RIS is deployed to help IoT devices' task offloading, there is a compromise for the UAV between the direct links and the links reflected by the RIS. By exploiting our proposed algorithm to adjust the phase shift of RIS, the reflected signals can be combined coherently to greatly improve the UAV’s received signal power. Therefore, the UAV tends to fly closer to the RIS rather than the IoT devices in order to fully utilize the channel gains brought by the RIS and increase the energy efficiency.

\begin{figure}[t]

\centering
\includegraphics[width =3.5in]{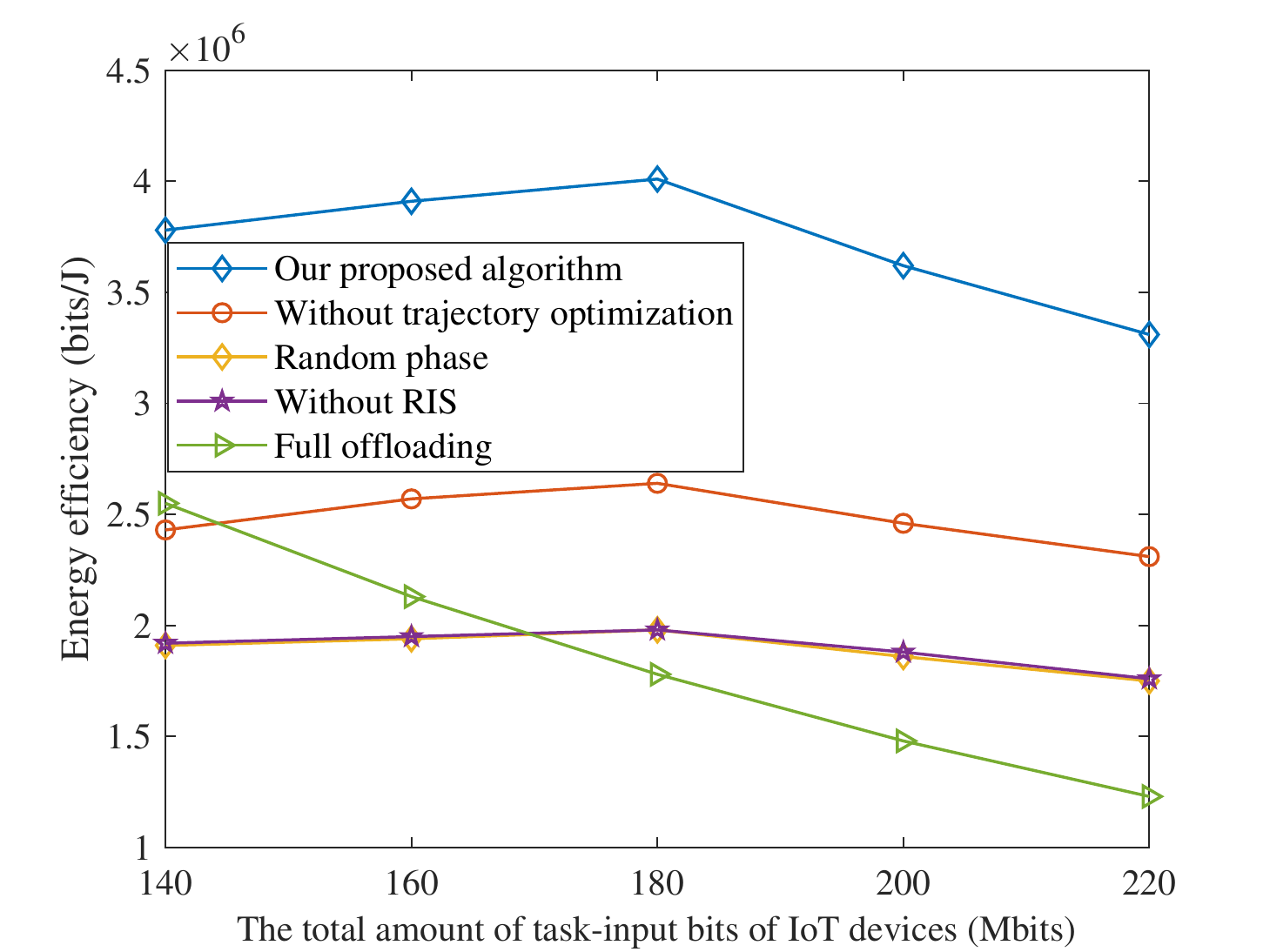}

\caption{Energy efficiency versus the total amount of task-input bits of IoT devices.}
\label{3}
\vspace{-0.1in}
\end{figure}

Fig. 4 shows the energy efficiency of the system versus the total amount of task-input bits of IoT devices, where our proposed energy efficiency maximization algorithm is compared with other four schemes: 1) Without trajectory optimization: The trajectory of UAV follows a straight line from the initial position to the final position. 2) Random phase: The phase shifts of RIS are randomly chosen from $\left[ {0,2\pi } \right]$. 3) Without RIS: The IoT devices offload their tasks without the aid of RIS. 4) Full offloading: The IoT devices are supposed to offload all task bits to the UAV for edge computing. It can be observed that with the deployment of RIS, our proposed algorithm can achieve a higher energy efficiency than the other schemes, since the transmit power, bit allocation, phase shift, and UAV trajectory are jointly optimized. Moreover, besides the full offloading scheme, it can be seen that the energy efficiency of the other schemes first increases and then decreases. The reasons is that the total offloading energy consumption of IoT devices can be expressed as an exponential function related with the offloading data rate according to Theorem 2. Since the increase rate of exponential function is faster than the linear function, with the increase of offloading data rate, the term \({{\sum\nolimits_{n = 1}^N {\sum\nolimits_{i = 1}^I {R_{{\pi _i}}^{{\rm{off}}}[n]t} } } \mathord{\left/
 {\vphantom {{\sum\nolimits_{n = 1}^N {\sum\nolimits_{i = 1}^I {R_{{\pi _i}}^{{\rm{off}}}[n]t} } } \Upsilon }} \right.
 \kern-\nulldelimiterspace} \Upsilon }\) in the energy efficiency first increases and then decreases, where \(\Upsilon \) is the right-hand-side (RHS) of (25). Thus, when the total amount of task-input bits increases, the energy efficiency first increases and then decreases. While for the full offloading scheme, the energy efficiency only shows a decrease trend due to the fact that the amount of offloading bits of IoT devices in the full offloading scheme is greatly larger than the other schemes. Besides, it can also be observed that the performance gain brought by the RIS over the scheme without RIS is negligible if the phase shifts are randomly chosen. This is because for the random phase scheme, the channel gain of the reflecting link is nearly equal to zero when those reflected signals via RIS are combined at the UAV. This result demonstrates the significance of phase shift optimization in the RIS-assisted UAV-enabled MEC system.

\begin{figure}[t]

\centering
\includegraphics[width =3.5in]{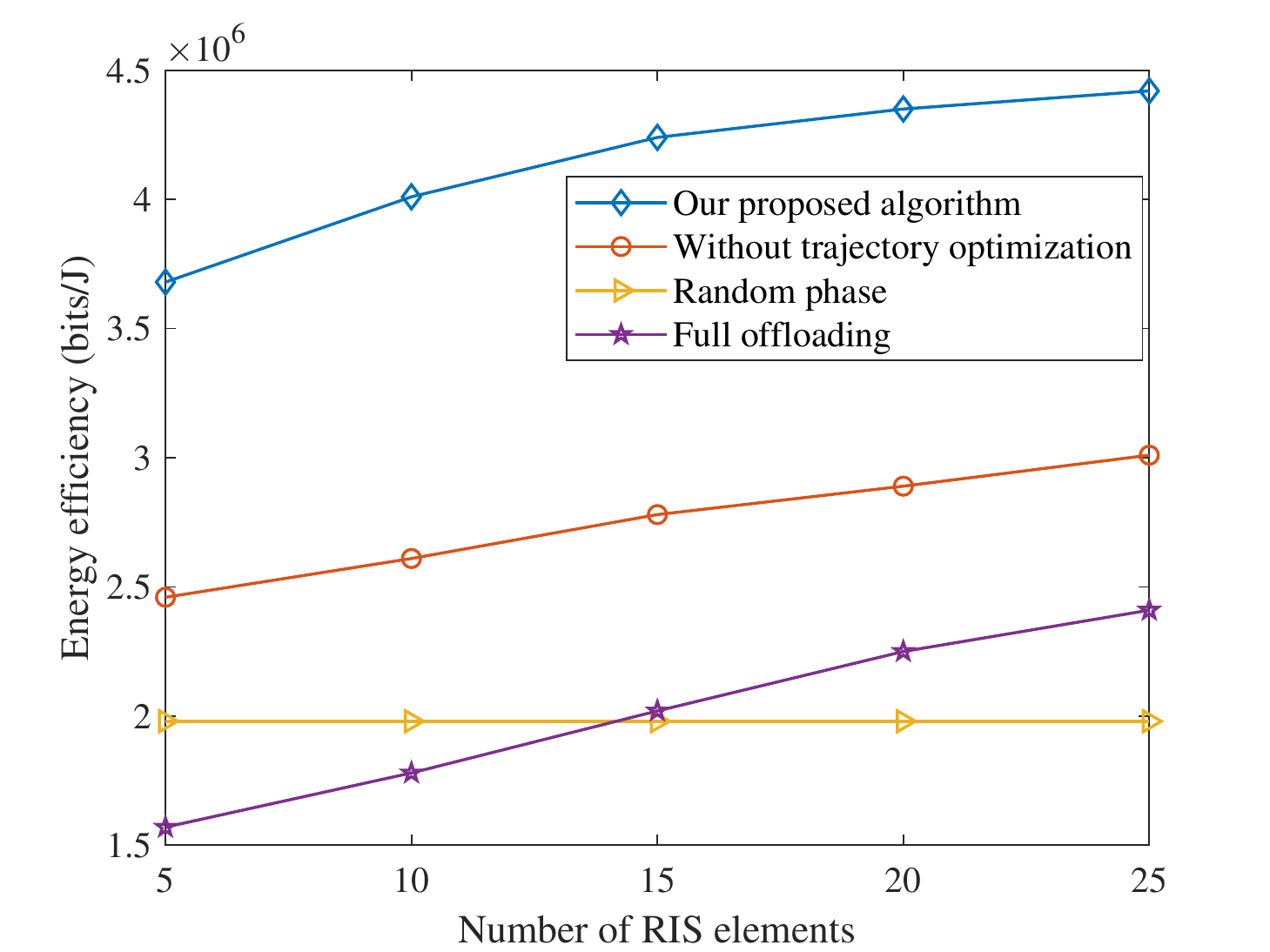}

\caption{Energy efficiency versus the number of RIS elements.}
\label{3}
\vspace{-0.1in}
\end{figure}

Fig. 5 presents the energy efficiency of the system versus the number of RIS elements, where the total amount of task-input bits is 180 Mbits. As expected, except the random phase scheme, the energy efficiency of all schemes increases as the number of RIS elements grows. The reason is that additional reflection elements will provide extra DoFs for designing more efficient phase shift strategy. Moreover, the performance gap between the proposed algorithm and the random phase scheme becomes larger as the number of RIS elements increase, which further verifies the necessity of phase shift optimization.


\begin{figure}[t]

\centering
\includegraphics[width =3.5in]{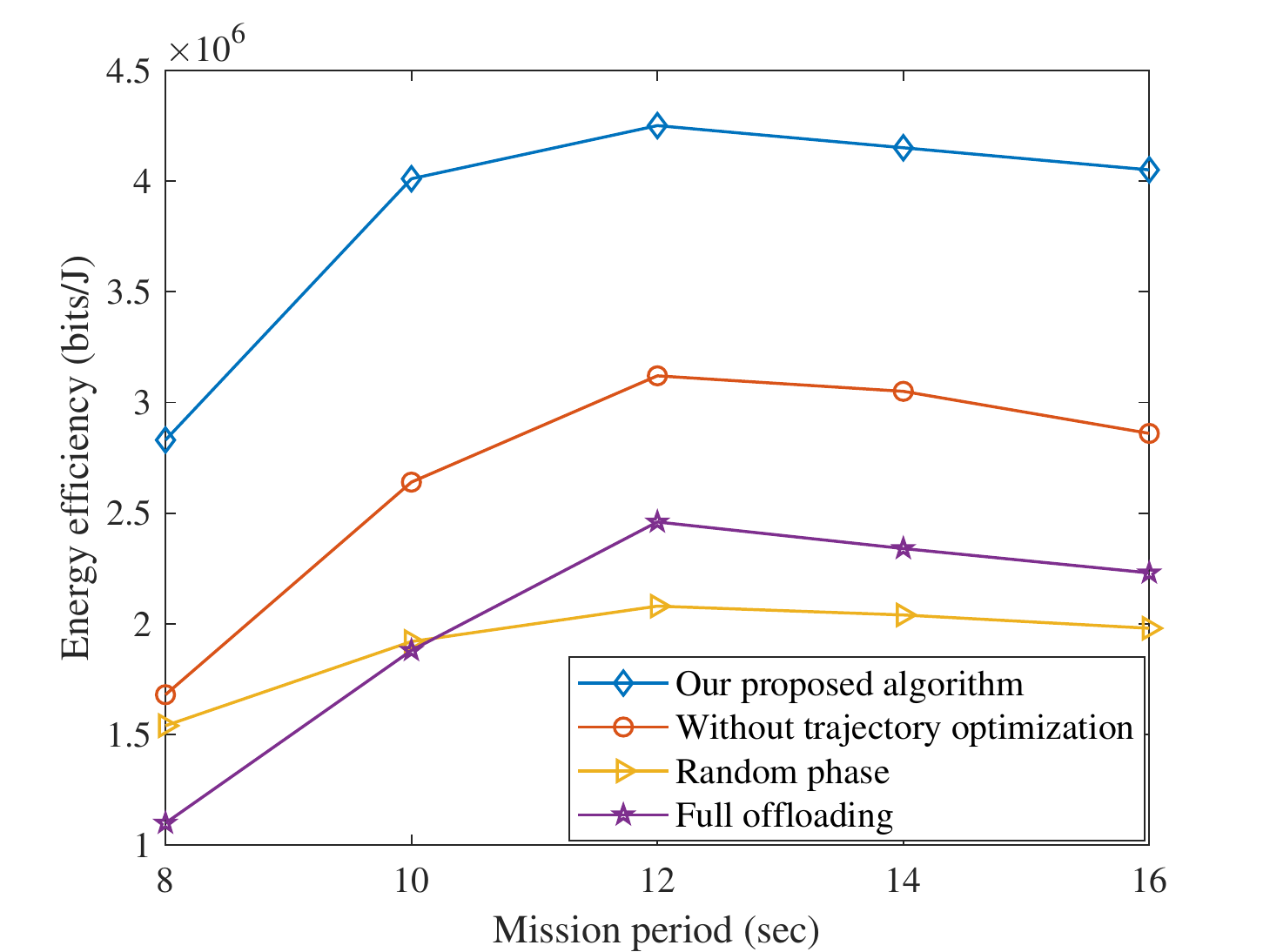}

\caption{Energy efficiency versus the mission period.}
\label{3}
\vspace{-0.1in}
\end{figure}

Fig. 6 shows the impact of the mission period $T$ on the energy efficiency. It is observed that our proposed algorithm can achieve higher energy efficiency compared with the other schemes. In addition, we also observe that the energy efficiency of all schemes increase with the mission period when the mission period is less than 12 sec. The reason is that a larger mission period enables the UAV to adjust its trajectory adequately. Thus, the channel conditions between the UAV and IoT devices can be effectively improved, which reduces the energy consumed by task offloading and accordingly improves the energy efficiency. Besides, with larger mission period, the task offloading and edge computing can be executed with longer time, which further reduce the energy consumed by computing. Nevertheless, with the further increase of mission period, the flying energy consumption of UAV continues to increase and dominates the total energy consumption. Thus, the energy efficiency gradual declines with the increase of mission period. Moreover, it can be seen that the increase of mission period brings more benefits for the full offloading scheme. This is because the amount of offloading bits for full offloading is larger than the other schemes and the IoT devices can take full advantages of the RIS. Therefore, when the IoT devices need to offload more tasks to the UAV-mounted MEC server, the energy efficiency can be greatly improved by properly increasing the mission period.

\begin{figure}[t]

\centering
\includegraphics[width =3.5in]{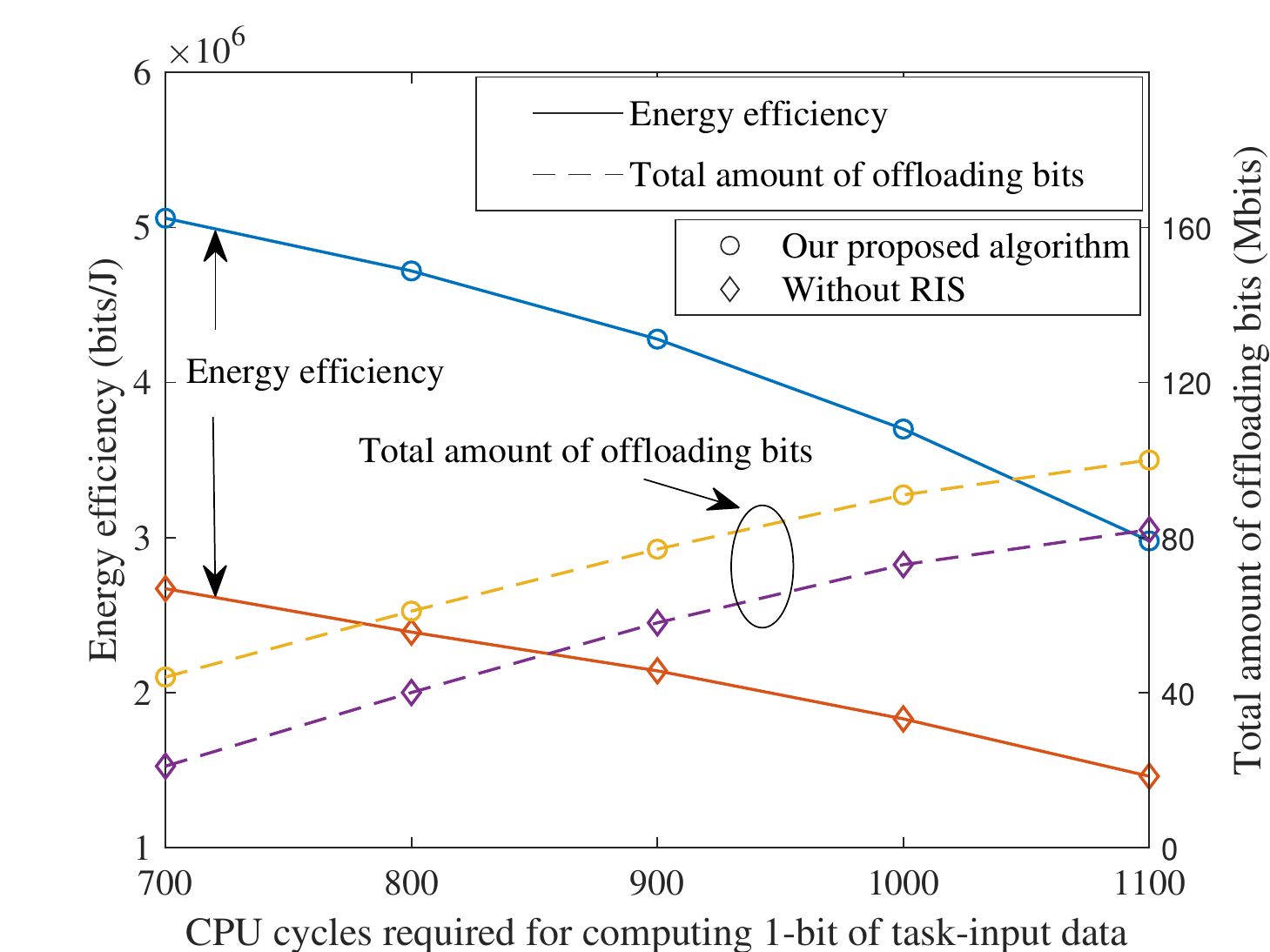}

\caption{Energy efficiency and the total amount of offloading bits versus the CPU cycles required for computing 1-bit of task-input data, $C_i$.}
\label{3}
\vspace{-0.1in}
\end{figure}


Fig. 7 illustrates the energy efficiency and the total amount of offloading bits versus the CPU cycles required for computing 1-bit of task-input data (i.e., $C_i$), where $T = 10\sec $ and the total amount of task-input bits of IoT devices is 200 Mbits. It can be seen that the energy efficiency decreases with the increase of $C_i$. This is because the increase of $C_i$ leads to more energy consumption for computing, and therefore results in the decrease of energy efficiency. Moreover, we observe that the amount of offloading bits increases when $C_i$ becomes larger. The reason is that the computing capacities of IoT devices are weaker than the UAV-mounted MEC server. In order to ensure the tasks can be completed within the mission period, the IoT devices have to offload more tasks to the UAV. In addition, compared with the scheme without RIS, we also observe that our proposed algorithm always offloads more task bits to the UAV in order to fully utilize the channel gains brought by the RIS to achieve higher energy efficiency.

\section{Conclusions}
In this paper, the RIS-assisted UAV-enabled MEC systems were investigated where the partial offloading scheme and the NOMA protocol were adopted for IoT devices’ task offloading. Aiming to maximize the energy efficiency, an iterative algorithm with a double-loop structure was proposed based on the Dinkelbach's method and BCD technique to jointly optimize the bit allocation, transmit power, phase shift, and UAV trajectory. Simulation results have shown that our proposed algorithm outperformed other baselines. It was also observed that with the aid of RIS, the energy efficiency can be greatly improved only when the phase shift was carefully designed, and the UAV tended to fly closer to the RIS to obtain a better channel condition, which was quite different from the UAV-enabled MEC without RIS. Besides, in order to ensure the tasks can be completed within the mission period, the IoT devices had to offload more task bits to the UAV when the CPU cycles required for computing 1-bit of task-input data became larger. Meanwhile, our proposed algorithm always offloaded more task bits to the UAV compared with the scheme without RIS.

\bibliographystyle{IEEEtran}
\bibliography{IEEEabrv,bib2014}

\end{sloppypar}
\end{document}